\documentclass{aa} 
\usepackage{txfonts}
\usepackage{ulem}
\usepackage{enumitem}
\usepackage{amsmath}
\usepackage{comment}

\usepackage{graphicx}
\usepackage{multirow} 
\usepackage{amssymb}
\usepackage{color}
\usepackage[switch]{lineno}
\usepackage{url}
\graphicspath{{../figures/}}
\usepackage{setspace}

\newcommand\U[1]{{\,\rm #1}}
\newcommand\E[1]{\times10^{#1}}
\newcommand\al{\alpha}
\newcommand\bt{\beta}
\newcommand\gm{\gamma}
\newcommand\Gm{\Gamma}
\newcommand\dl{\delta}
\newcommand\tht{\theta}

\newcommand\sg{\sigma}
\newcommand\om{\omega}
\newcommand\upi{\pi}

\newcommand\rs[1]{_\mathrm{#1}}
\newcommand\elm{m\rs{e}}
\newcommand\omc{\om\rs{c}}
\newcommand\gmcut{\gm\rs{cut}}
\newcommand\xcut{x\rs{cut}}
\newcommand\omcut{\om\rs{cut}}
\newcommand\nubr{\nu\rs{br}}
\newcommand\Wo{W\rs{o}}
\newcommand\Pio{\Pi\rs{o}}
\newcommand\Fn{F\rs{n}}
\newcommand\Gn{G\rs{n}}
\newcommand\sgeff{\sg\rs{eff}}
\newcommand\fanis{f\rs{an}}

\newcommand\Bbar{\bar B}
\newcommand\XFappr{X_F^{\rm appr}}
\newcommand\XGappr{X_G^{\rm appr}}

\newcommand\sgpar{\sigma_\parallel}
\newcommand\sgperp{\sigma_\perp}
\newcommand\sgupar{\sigma\rs{u,\parallel}}
\newcommand\sguperp{\sigma\rs{u,\perp}}
\newcommand\sggpar{\sigma\rs{g,\parallel}}
\newcommand\sggperp{\sigma\rs{g,\perp}}

\newcommand\Bbarupar{{\bar B}\rs{u,\parallel}}
\newcommand\Bbaruperp{{\bar B}\rs{u,\perp}}
\newcommand\Bbaru{{\bar B}\rs{u}}
\newcommand\sgu{\sigma\rs{u}}
\newcommand\Bbardpar{{\bar B}\rs{d,\parallel}}
\newcommand\Bbardperp{{\bar B}\rs{d,\perp}}
\newcommand\sgdpar{\sigma\rs{d,\parallel}}
\newcommand\sgdperp{\sigma\rs{d,\perp}}
\newcommand\thtu{\theta\rs{u}}
\newcommand\thtd{\theta\rs{d}}

\newcommand\Bperp{B_\perp}
\newcommand\Pperp{P_\perp}
\newcommand\Ppar{P_\parallel}
\newcommand{\defeq}{\mathrel{\hat{=}}}

\usepackage{natbib,twoopt}
\usepackage[breaklinks=true]{hyperref} %% to avoid \citeads line fills
\bibpunct{(}{)}{;}{a}{}{,}             %% natbib format for A&A and ApJ
\makeatletter
  \newcommandtwoopt{\citeads}[3][][]{\href{http://adsabs.harvard.edu/abs/#3}%
    {\def\hyper@linkstart##1##2{}%
     \let\hyper@linkend\@empty\citealp[#1][#2]{#3}}}
  \newcommandtwoopt{\citepads}[3][][]{\href{http://adsabs.harvard.edu/abs/#3}%
    {\def\hyper@linkstart##1##2{}%
     \let\hyper@linkend\@empty\citep[#1][#2]{#3}}}
  \newcommandtwoopt{\citetads}[3][][]{\href{http://adsabs.harvard.edu/abs/#3}%
    {\def\hyper@linkstart##1##2{}%
     \let\hyper@linkend\@empty\citet[#1][#2]{#3}}}
  \newcommandtwoopt{\citeyearads}[3][][]%
    {\href{http://adsabs.harvard.edu/abs/#3}
    {\def\hyper@linkstart##1##2{}%
     \let\hyper@linkend\@empty\citeyear[#1][#2]{#3}}}
\makeatother
\title{Synchrotron polarization with a partially random magnetic field:\\
general approach, and application to X-ray polarization from SNRs}
\titlerunning{Synchrotron polarization with a partially random magnetic field}
\authorrunning{R.Bandiera \& O.Petruk}

\author{Rino Bandiera\inst{1}, Oleh Petruk\inst{2,3}}
\institute{INAF - Osservatorio Astrofisico di Arcetri, Largo E. Fermi 5, I-50125 Firenze, Italy
\and
INAF - Osservatorio Astronomico di Palermo, Piazza del Parlamento 1, I-90134 Palermo, Italy
\and
Institute for Applied Problems in Mechanics and Mathematics, National Academy of Sciences of Ukraine, Naukova St. 3-b, 79060 Lviv, Ukraine\\
\email{rino.bandiera@inaf.it, oleh.petruk@inaf.it}
}
\abstract
{Diagnostics based on the polarization properties of the synchrotron emission can provide precious information on both the ordered structure and the random level of the magnetic field. While this issue has been already analysed in the radio band, the polarization data recently obtained by the mission IXPE have shown the need to extend this analysis to the X-ray band.}
{While our immediate target are young supernova remnants, the scope of this analysis is wider. We aim at extending the analysis to particle energy distributions more complex than a power law, as well as to investigate a wider range of cases involving a composition of ordered and random magnetic fields.}
{Since only in a limited number of cases an analytical approach is possible, we have devised to this purpose an optimised numerical scheme, and we have directly used it to investigate particle energy distributions in the form of a power law with an exponential, or super-exponential cutoff. We have also considered a general combination of a ordered field plus an anisotropic random component.}
{We have shown that the previously derived analytic formulae, valid for power-law distributions, may be good approximations of the  polarization degree also in the more general case with a cutoff, as typically seen in X-rays. We have explicitly analysed the young supernova remnants SN~1006, Tycho and Cas A. In particular, for SN~1006, we have proved the consistency between the radio and X-ray polarization degrees, favouring the case of predominantly random field, with an anisotropic distribution. In addition, for the power-law case, we have investigated the effect of a compression on both ordered and random magnetic field components, aimed at describing the mid-age radio supernova remnants.}
{This work allows a more efficient exploitation of radio and X-ray measurements of the synchrotron polarization, and is addressed to present observations with IXPE as well as to future projects.}

\begin{document}
%
%\flushbottom
\maketitle
\thispagestyle{empty}
%

%%%%%%%%%%%%%%%%%%%%%%%%%%%%%%%%%%%%%%%%%%%%%%%%%
\section{Introduction}
\label{sec:intro}

Relativistic particles spiralling in a background magnetic field (MF) emit synchrotron radiation.
In the presence of a homogeneous MF, this radiation is characterised by a very high polarization degree ($\Pi$).
However, homogeneous fields rarely occur in real cases: in the majority of cases, instead, fields can be disordered to some extent; in addition, large scale structures may cause a mix of different MF conditions along the line of sight; and even limitations in the instrumental angular resolution may have as effect a superposition of differently oriented MFs.

From the analysis of observations of synchrotron emission one can derive some properties of the MF and the energy distribution of relativistic particles.
In the presence of an inhomogeneous MF, a more careful modelling is required for this derivation, and this is the primary goal of the present work.
Here we will mainly focus to the case of Supernova Remnants (SNRs), but many of the results presented can be applied also to other sources.

In a former paper \citep[][hereafter Paper~I]{2016MNRAS.459..178B}, we had thoroughly discussed the behaviour of the synchrotron emission and of its $\Pi$, in the presence of a MF in which a homogeneous component is (vectorially) combined with a random one, while the relativistic particles are assumed to be described by a power-law energy distribution.
This study led to analytic, although rather complex, formulae that apply for instance to the case of radio emission from SNRs, where the synchrotron emission is usually very well approximated by a power law. In particular, we have used that approach to analyse the general properties of polarization images of Sedov SNRs in a uniform interstellar medium \citep{2017MNRAS.470.1156P}.

The observational possibilities have been extended dramatically with the launch of the IXPE mission \citep[see e.g.][]{2022HEAD...1930101W}, which for the first time has provided good quality X-ray polarization maps of several sources.
In the case of Pulsar Wind Nebulae also the X-ray spectra are power laws, so that our original results can be safely applied, as actually done for the Crab Nebula \citep{2023NatAs...7..602B}, Vela \citep{2023ApJ...959L...2L}, and G0.13--0.11 \citep{2023arXiv231204421C}.
Instead, synchrotron X-ray spectra of shell-type SNRs are characterised by a cutoff.
Our formulae have been also used to discuss the polarization degrees in Cas A, Tycho SNR and SN 1006
\citep{2023ApJ...945...52F, 2023ApJ...957...55Z} but, without an extension of our theoretical treatment to spectra with a cutoff, the level of accuracy of those estimates could not be assessed.

While an exact analytic treatment to the more general problem seems to be impossible, we have devised a new numerical treatment that allows us to 
extend the analysis of Paper I and to treat with a high level of accuracy cases in which the particle distribution is more complex than a power law.

The plan of the paper is as follows.
In Sect.~\ref{sec:method} we review some basic concepts from Paper~I, and describe the numerical approach we have developed here;
Sections~\ref{sec:calciso} and \ref{sec:calcaniso} consider, respectively, the case of an isotropic random MF combined with a homogeneous one, and that of a purely random anisotropic MF, and for each of them show the results for particle distributions well described by a power law times an exponential, or super-exponential, cutoff. 
In Sect.~\ref{sec:observable} we apply our techniques to observable quantities such as the polarization degree, the spectral index, the spectral curvature, and we discuss the relations between them and with the properties of the emitting system. We also discuss with which accuracy our analytic formulae from Paper I (derived for the power-law electron spectrum) could be used in a more general case of particle distribution with a high-energy cut-off.
Section~\ref{sec:youngSNRs} discusses some young SNRs, and in particular SN~1006, for which the mapping of the polarization degree in X-rays with IXPE is particularly detailed, and allows a close comparison with the polarization in radio.
Section~\ref{sec:homo+anis} investigates the case of the ordered + anisotropic random MF components.
Section~\ref{sec:conclusions} concludes.

%%%%%%%%%%%%%%%%%%%%%%%%%%%%%%%%%%%%%%%%%%%%%%%%%
\section{The method of calculation}
\label{sec:method}

We will first review some basic formulae of the classical theory of the synchrotron emission.
As a reference for the formulae and notation we have used the book by \citet{1986rpa..book.....R}.

Let us consider the average emission from a single particle with a given Lorentz factor $\gm$.
In this case, the synchrotron power emitted per unit frequency \footnote{for the frequency, we often use the variable $\om=2\upi\nu$, so that $P(\nu)=2\upi\,P(\om)$} can be computed as the sum of two polarised components, respectively perpendicular ($\Pperp$) and parallel ($\Ppar$) to the direction of the projected MF (labelled $\Bperp$, being perpendicular to the Line of Sight, LoS):
%%%
\begin{eqnarray}
    \Pperp(\om)&=&\frac{\sqrt{3}\,e^3\Bperp}{4\upi\,\elm\,c^2}\Big(F(x)+G(x)\Big); \\
    \Ppar(\om)&=&\frac{\sqrt{3}\,e^3\Bperp}{4\upi\,\elm\,c^2}\Big(F(x)-G(x)\Big),
\end{eqnarray}
%%%
where:
%%%
\begin{eqnarray}
    F(x)&=&x\int_x^\infty{K_{5/3}(z)\,dz};
    \label{eq:Fdef}\\
    G(x)&=&x\,K_{2/3}(x),
    \label{eq:Gdef}
\end{eqnarray}
%%%
with $K_n(z)$ being a modified Bessel function of the second kind, while the variable $x$ is defined as:
%%%
\begin{equation}
    x=\frac{\om}{\omc}=\frac{2}{3}\frac{\elm c}{e}\frac{\om}{\Bperp\gm^2}\defeq\frac{2 K}{\Bperp\gm^2},
    \label{eq:xdef}
\end{equation}
%%%
(the last equation representing a definition for $K$); $\omc$ is called ``critical frequency''.

Let us first consider an orientation of the local axes (namely in the volume element under consideration) such that the unit vector $\hat x'$ is perpendicular to $\Bperp$ while $\hat y'$ is parallel to it.
With respect to these axes, the Stokes parameters ${\cal I}'$ (total flux) and ${\cal Q}'$ (linear polarization) read:
%%%
\begin{eqnarray}
    {\cal I}'(\om)&\!\!\!=\!\!\!&\frac{\Pperp+\Ppar}{4\upi}\,=\,\frac{\sqrt{3}\,e^3}{8\upi^2\,\elm\,c^2}\Bperp F(x)\,\defeq\,H\,\Bperp F(x)
    \label{eq:Iprime}\\
    {\cal Q}'(\om)&\!\!\!=\!\!\!&\frac{\Pperp-\Ppar}{4\upi}\,=\,\frac{\sqrt{3}\,e^3}{8\upi^2\,\elm\,c^2}\Bperp G(x)\,\defeq\,H\,\Bperp G(x),
    \label{eq:Qprime}
\end{eqnarray}
%%%
while the other two Stokes parameters are vanishing, ${\cal U}'$ due to the chosen orientation of axes, ${\cal V}'$ due to the properties of the synchrotron radiation.
In the following, while still considering the MF perpendicular to the LoS, we will no longer use the symbol $\Bperp$, but simply $B$ (but still having in mind that it is a 2-D vector).

We also introduce a fixed reference coordinate system, $x$--$y$, which in the following we will refer to as the ``observer's coordinate system''. 
Let us now represent $B$ as the composition of the homogeneous MF $\bar B$ that, without loss of generality, we assume directed along the $y$ axis, and of a random MF.
Let us label as $B_x$ and $B_y$ the components of the combined field, in the observer's coordinate system, and ${\cal I}$, ${\cal Q}$, and ${\cal U}$ are Stokes parameters in this system (in the case of a symmetric distribution for the random MF we expect to have ${\cal U}=0$).

The $x'$--$y'$ coordinate system will be wiggling all the time, so that in order to derive time averaged quantities we always have to account for a rotation on a suitable angle $\chi$ to recover the same orientation of the fixed $x$--$y$ coordinate system.
This rotation has no effect on ${\cal I}$ (${\cal I}={\cal I}'$), while the Stokes parameters ${\cal Q}$ and ${\cal U}$ transform according to ${\cal Q}=\cos(2\,\chi)\,{\cal Q}'$ and ${\cal U}=\sin(2\,\chi)\,{\cal Q}'$, where
%%%
\begin{eqnarray}
 \label{eq:costwochidef}
    \cos(2\,\chi)&=&\frac{B_y^2-B_x^2}{B^2};    \\
 \label{eq:sintwochidef}
    \sin(2\,\chi)&=&-\frac{2B_xB_y}{B^2}
\end{eqnarray}
%%%
and the angle $\chi$ between the axis $y$ and the vector ${B}$ is measured anticlockwise by the observer.\footnote{This is the same angle as between the axis $x$ an the $E$-polarization direction. It is measured in the same direction as the angle of rotation due to the Faraday effect.}
With the use of the above formulae, in Paper~I we have first integrated the Stokes parameters over the probability distribution of the random MF components.
For this, we have explicitly considered two cases: 1. a homogeneneous field plus an isotropic random component; 2. an anisotropic random MF, plus a negligible homogeneous MF.

A clarification is needed about this point. By using this approach, namely by integrating over a probability distribution for the random MF, we implicitly assume that virtually all possible realizations are reached, with frequencies very close to their assigned probabilities. In principle, this is only valid for infinite systems. On the contrary, in the case of only a few realizations along the LoS, our method gives only the statistical average, while in the various cases, taken individually, one could measure a more or less relevant dispersion of values. Such dispersion may be relevant in the case of a developed turbulence with an injection length scale comparable with the length of the LoS, while it should be negligible in the case of an injection length scale much smaller than the LoS. 

The treatment of real turbulence is beyond the scope of the present work, and it will be discussed in a forthcoming paper. Here we limit our description to a ``random MF'', namely fully characterised by its probability distribution function. Anyway, our results will hold also in the case of a finite although rather large number of statistically independent cells along the LoS. In this, the size of cells that could possibly create measurable effects is so large that the MF conditions in them are frozen during the time of observation; while the only effect of a finite exposition time is to account for the uncertainties during the model fitting. 
Indeed, the longest exposure times, with IXPE, do not exceed one month, while random motions should be much slower than 1000 km/s. Even in this very extreme case, the size of a cell in which magnetic field changes during an observation run should be smaller than $10^{-4}\U{pc}$ (compared to a size of SNR $\sim 1\U{pc}$); in addition, in the case of real turbulence the power spectrum of the velocity fluctuations is smaller at smaller scales.
Moreover, in the case of lower instrumental resolution the contributions of different LoSs will add up, so that the effective number of the independent cells will be even larger, and consequently the level of possible fluctuations around the mean value in the distribution function will be even smaller.

In a second stage, we have then integrated over the particles energy distribution, a power law in that case.
Thanks to simplifications for the special case of a power-law distribution, in Paper~I we have obtained analytic formulae, although involving special functions.

By releasing the condition of a power-law particle distribution, the problem becomes by far more complex, so that analytic formulae cannot be longer obtained.
For this reason a numerical approach is required, but we have taken anyway maximum advantage of some analytical results to speed up dramatically the computations.
In this new approach we have found more convenient to proceed in the opposite order with respect to Paper~I, by performing first the integration over the particles distribution, while later on the average over the MF fluctuations.
We have considered the following family of particle energy distributions:
%%%
\begin{equation}
    n(\gm)=A\gm^{-s}\exp\left(-(\gm/\gmcut)^\bt\,\right),
    \label{eq:ndistrib}
\end{equation}
%%%
with positive $s$ and $\bt$; $\gmcut$ indicates the position of the cutoff in the particle energy distribution.
Therefore, for ${\cal I}'$ integrated over the energy distribution, we can write as follows:
%%%
\begin{eqnarray}
    {\cal I}'&\!\!\!\!=\!\!\!\!&AHB\int_0^\infty\!\!\!{F(x)\,\gm^{-s}\exp\left(-(\gm/\gmcut)^\bt\,\right)\,d\gm}     \nonumber\\
    &\!\!\!\!=\!\!\!\!&\frac{AHB^{(s+1)/2}}{4K^{(s-1)/2}}\int_0^\infty{F(x)\,\left(\frac{x}{2}\right)^{(s-3)/2}\!\!\!\!\!\!\!\exp\left(-\left(\frac{x}{\xcut}\right)^{-\bt/2}\,\right)\,dx},\qquad\quad
    \label{eq:Iprimedef}
\end{eqnarray}
%%%
where:
%%%
\begin{equation}
    \xcut=\frac{2K}{B\,\gmcut^2}=\frac{2}{3}\frac{\elm\,c}{e}\frac{\om}{B\,\gmcut^2}\defeq\frac{\om}{\omcut}.
\end{equation}
%%%
By recalling that the peak of synchrotron emission for a given Lorentz factor is at $\om=0.29\,\omc$ (this quantity is also conventionally adopted in the ``monochromatic'' approximation of the synchrotron emission), in order to conform to this standard approach let us define the ``break frequency'' as:
%%%
\begin{equation}
    \nubr=\frac{0.29}{2\upi}\omcut.
\end{equation}
%%%
In a similar way to what obtained for ${\cal I}'$, we also get:
%%%
\begin{equation}
    {\cal Q}'=\frac{AHB^{(s+1)/2}}{4K^{(s-1)/2}}\int_0^\infty{G(x)\,\left(\frac{x}{2}\right)^{(s-3)/2}\exp\left(-\left(\frac{x}{\xcut}\right)^{-\bt/2}\,\right)\,dx}.
    \label{eq:Qprimedef}
\end{equation}
%%%
In the pure power-law case, the standard result is recovered:
%%%
\begin{eqnarray}
    {\cal I}'\rs{PL}&=&\frac{s+7/3}{s+1}\Wo B^{(s+1)/2};    
    \label{eq:IPLprime}\\
    {\cal Q}'\rs{PL}&=&\qquad\quad\,\Wo B^{(s+1)/2},
    \label{eq:QPLprime}
\end{eqnarray}
%%%
where:
%%%
\begin{equation}
    \Wo=\frac{AH}{4K^{(s-1)/2}}\Gm\left(\frac{s}{4}+\frac{7}{12}\right)\Gm\left(\frac{s}{4}-\frac{1}{12}\right).
    \label{eq:W0def}
\end{equation}
%%%
With the more general form of energy distribution, as from Eq.~\ref{eq:ndistrib}, a new numerical calculation must be performed for any choice of $s$ and $\bt$.
It is already rather complex in the case of a homogeneous MF, but it becomes excessively heavy in the presence of also a random MF component.
A brute force approach to this problem would involve the calculation of many random instances for the MF, according to a given distribution function of the random MF component, and then to average among all these cases.
For each actualization of the $B_x$ and $B_y$ field components we must first derive the total MF $B$, after which to compute the quantities ${\cal I}'$ (equal to ${\cal I}$) and ${\cal Q}'$ with the use of Eqs.~\ref{eq:Iprimedef} and \ref{eq:Qprimedef} (this is the most cumbersome part, because it involves the calculation of rather heavy integrals), and then obtain ${\cal Q}$ and ${\cal U}$ with the use of Eqs.~\ref{eq:costwochidef} and \ref{eq:sintwochidef}.

The Stokes parameters are finally obtained as the averages over all these instances.
Unfortunately, this approach converges rather slowly, due to the fact that the uncertainty on the average is proportional to the inverse square root of the number of cases used.
However, there is a way to significantly speed up the calculations, without any substantial loss of accuracy.
One may note that the formulae for ${\cal I}'$ and ${\cal Q}'$, the quantities whose evaluation absorbs most of the time, do not depend on both $B_x$ and $B_y$, but only on the MF module $B$.
So, an array of their values could be computed first, and then interpolations could be used for each actualization of the MF.

For a given probability distribution ${\cal P}(B_x,B_y)$, of the components of the total (transverse) MF $B$, the average Stokes parameters are then evaluated in the observer's frame as:
%%%
\begin{eqnarray}
    \label{eq:Istokesdef}
    \langle{\cal I}\rangle&=&\int {\cal I'}(B)\,{\cal P}(B_x,B_y)\,dB_xdB_y;\\
    \langle{\cal Q}\rangle&=&\int {\cos(2\chi)\,\cal Q'}(B)\,{\cal P}(B_x,B_y)\,dB_xdB_y,\\
    \langle{\cal U}\rangle&=&\int {\sin(2\chi)  \,\cal Q'}(B)\,{\cal P}(B_x,B_y)\,dB_xdB_y
\end{eqnarray}
%%%
(note that the formulae above assume only variations of the MF, without changes in the density of the emitting particles). Assuming a symmetric distribution ${\cal P}(B_x,B_y)$ with respect to $B_x$, the parameter $\langle{\cal U}\rangle$ vanishes, and the polarization degree is $\Pi=\langle{\cal Q}\rangle/\langle{\cal I}\rangle$.
We may introduce the functions $X_F={\cal I}'/{\cal I}'\rs{PL}$ and $X_G={\cal Q}'/{\cal Q}'\rs{PL}$, namely:
%%%
\begin{eqnarray}
    X_F\!\!\!\!\!\!&=&\!\!\!\!\!\!
    \frac{(s+1)/(s+7/3)}{\Gm\left(\frac{s}{4}+\frac{7}{12}\right)\Gm\left(\frac{s}{4}-\frac{1}{12}\right)}\!\int_0^\infty{\!\!\!\!\!F(x)\,\left(\frac{x}{2}\right)^{(s-3)/2}e^{-(x/\xcut)^{-\bt/2}}\!dx};\quad
    \label{eq:XFdef}
    \\
    X_G\!\!\!\!\!\!&=&\!\!\!\!\!\!
    \frac{1}{\Gm\left(\frac{s}{4}+\frac{7}{12}\right)\Gm\left(\frac{s}{4}-\frac{1}{12}\right)}\!\int_0^\infty{\!\!\!\!\!G(x)\,\left(\frac{x}{2}\right)^{(s-3)/2}e^{-(x/\xcut)^{-\bt/2}}\!dx}.\quad
    \label{eq:XGdef}
\end{eqnarray}
%%%
For a given choice of $s$ and $\bt$, $X_F$ and $X_G$ are functions of $\xcut$ only. 
In addition, both functions, defined in this way, approach asymptotically 1 for $\xcut\rightarrow0$
(see Appendix~\ref{app:fits2intgFG} for more details).
Therefore, the above formulae translate into:
%%%
\begin{eqnarray}
    \label{eqs:aveI}
    \langle{\cal I}\rangle\!\!\!\!\!\!&=&\!\!\!\!\!\!\frac{\Wo}{\Pio}\!\!\int{\!\!B^{(s+1)/2}X_F\left(
    %\frac{2}{3}\frac{\elm c}{e}\!\frac{\om}{B\gmcut^2}
    \xcut
    \right)}\,{\cal P}(B_x,B_y)\,dB_x\,dB_y;\quad\,  \\
    \label{eqs:aveQ}
    \langle{\cal Q}\rangle\!\!\!\!\!\!&=&\!\!\!\!\!\!\Wo\!\!\!\int{\!\!\frac{B_y^2-B_x^2}{B^2}B^{(s+1)/2}X_G\left(
    %\frac{2}{3}\frac{\elm c}{e}\!\frac{\om}{B\gmcut^2}
    \xcut
    \right)}\,{\cal P}(B_x,B_y)\,dB_x\,dB_y\quad\,
\end{eqnarray}
%%%
where
%%%
\begin{equation}
    \label{eqs:Pio}
    \Pio=(s+1)/(s+7/3)
\end{equation}
%%%
is the polarization degree for a homogeneous MF.

Let us now assume that the components of the total (transverse) MF $B$ are well described by a Gaussian probability distribution:
%%%
\begin{equation}
    {\cal P}(B_x,B_y)=\frac{1}{2\upi\,\sg_x\sg_y}\exp\left(-\frac{B_x^2}{2\,\sg_x^2}-\frac{(B_y-\Bbar)^2}{2\,\sg_y^2}
    \right).
\end{equation}
%%%
The transition to the homogeneous case (when the random MF component tends to zero and the only ordered component is important) is provided by the limits $\sigma_x\rightarrow 0$ and $\sigma_y\rightarrow 0$, in which case ${\cal P}(B_x,B_y)=\dl\left(B_x\right)\cdot\dl\left(B_y-\Bbar\right)$.

In the next two sections we will present and discuss the results, applied respectively to the isotropic random case ($\sg_x=\sg_y\defeq\sg$), and to the anisotropic case with a vanishing ordered field ($\Bbar=0$); in other terms, we will study the two cases corresponding to those treated in our Paper 1.
In Sect.~\ref{sec:homo+anis} a more general anisotropic plus $\Bbar\neq0$ problem is considered.

%%%%%%%%%%%%%%%%%%%%%%%%%%%%%%%%%%%%%%%%%%%%%%%%%
\subsection{Results for an isotropic random MF}
\label{sec:calciso}
%%%%%%%%%%%%%%%%%%%%%%%%%%%%%%%%%%%%%%%%%%%%%%%%%
\begin{figure} 
\centering
	\includegraphics[width=.47\textwidth]{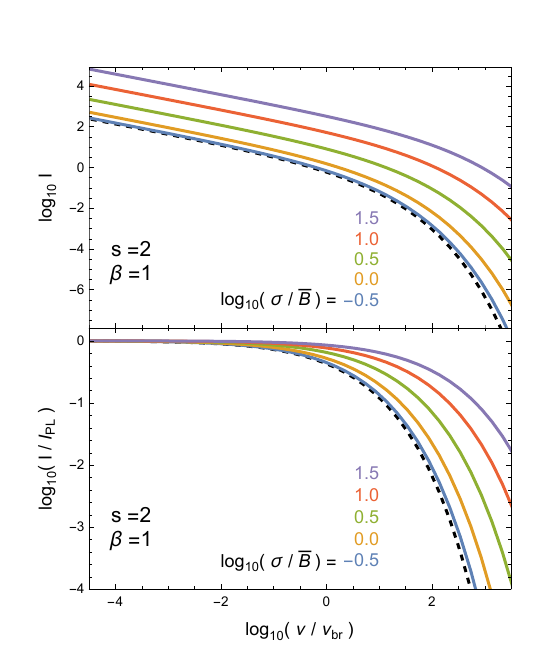}
        \caption{Upper panel: spectra of ${\cal I}$, for a fixed choice of $s$ ($=2$) and $\bt$ ($=1$), and different choices of the level of the random MF, here isotropic and characterised by the standard deviation $\sg$. ${\cal I}$ is normalised by taking $A=H=K=\xcut=\Bbar=1$ in Eqs.~\ref{eq:Iprimedef} and \ref{eq:Istokesdef}.
        The dashed line represents the case without random MF; while the other color lines refer to values of $\log_{10}(\sg/\Bbar)$ ranging between $-0.5$ and $+1.5$, in steps of $0.5$. Lower panel: same as above, but now scaled with the pure power-law solution from Paper~I.
        }
\label{fig:curvesIs2bt1iso}	
\end{figure}
%%%%%%%%%%%%%%%%%%%%%%%%%%%%%%%%%%%%%%%%%%%%%%%%%
\begin{figure}
\centering
\includegraphics[width=.47\textwidth]{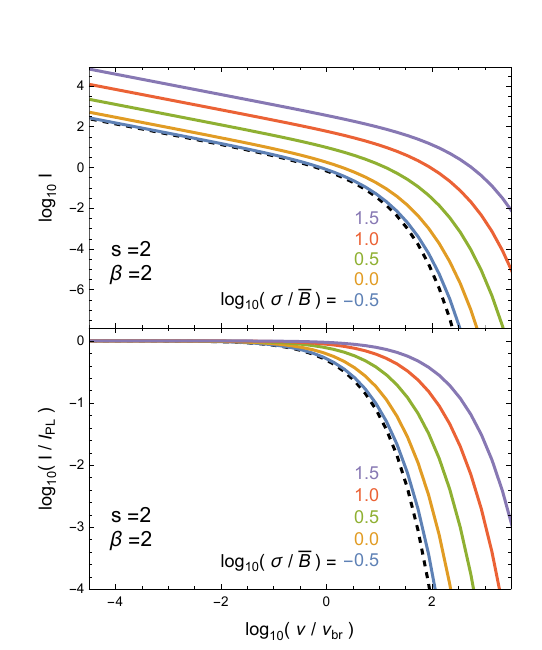}
        \caption{Same as Fig.~\ref{fig:curvesIs2bt1iso}, but now for $\bt=2$.
        }
\label{fig:curvesIs2bt2iso}	
\end{figure}
%%%%%%%%%%%%%%%%%%%%%%%%%%%%%%%%%%%%%%%%%%%%%%%%%
\begin{figure} 
\centering
    \includegraphics[width=.47\textwidth]{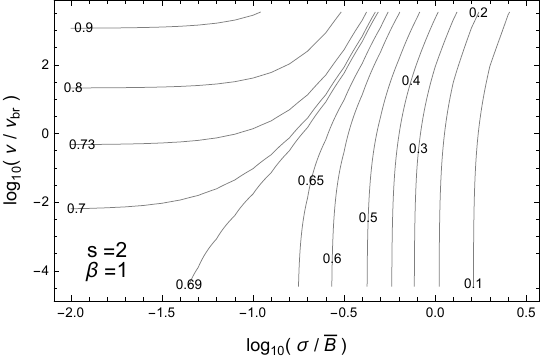}
        \caption{Iso-levels representation of the $\Pi$ dependence on the level of the (isotropic) random MF and on the frequency, for cases with $s=2$, $\bt=1$. The homogeneous case is approached on the left edge of the figure, while on the bottom edge we have a good approximation of the power-law case.}
\label{fig:figPDs2bt1iso}	
\end{figure}
%%%%%%%%%%%%%%%%%%%%%%%%%%%%%%%%%%%%%%%%%%%%%%%%%
\begin{figure} 
\centering
	\includegraphics[width=.47\textwidth]{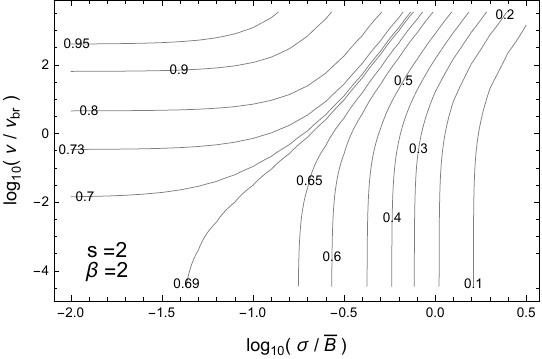}
        \caption{Same as Fig.~\ref{fig:figPDs2bt1iso}, for cases with $\bt=2$.
        }
\label{fig:figPDs2bt2iso}	
\end{figure}
%%%%%%%%%%%%%%%%%%%%%%%%%%%%%%%%%%%%%%%%%%%%%%%%%
Here we will focus our analysis to the case $s=2$, which is the value found theoretically in the case of a (non radiative, non cosmic-ray modified) strong shock, and that also agrees with the average $-0.5$ spectral index measured in radio for the synchrotron emission from SNRs.
In Appendix \ref{app:fits2intgFG} some considerations will also be presented for values of $s$ close to this reference value.
Let us consider two different values for $\bt$, namely $\bt=1$ and $\bt=2$, which should well approximate the cases in which the cutoff in the energy distribution of the accelerated particles is due, respectively, to the finite time of the acceleration process (like to the SNR age), or to radiative losses \citep[see also][]{2007A&A...465..695Z}.

The results on the spectra of ${\cal I}$ are summarised in Figs.~\ref{fig:curvesIs2bt1iso} and \ref{fig:curvesIs2bt2iso}, for the case of an isotropic random MF (i.e. $\sg_x=\sg_y\defeq\sg$).
The trends look qualitatively as expected: for frequency higher than the cutoff frequency the emission become dramatically lower than the power-law extrapolation, and this effect is stronger at larger values of $\bt$; the emission is also larger for larger values of $\sg$, since the combined MF is generally \rm{higher}.
Of course, in real cases only the beginning of this spectral bending could be tested, because further on the emission would be too weak to be detectable.

Since the absolute value of the emission strongly depends on quantities (like the particles density and the MF intensity) that are poorly known, in practice the most directly testable quantity is $\Pi$.
Figures~\ref{fig:figPDs2bt1iso} and \ref{fig:figPDs2bt2iso} give, again in the case of an isotropic random MF and the parameters setting described above, the dependence of $\Pi$ on $\sg$ and $\nu$.
The bottom side of the plots (low $\nu/\nubr$) represent cases still in the power-law region of the spectrum; while their left-hand side represents cases with a very low random field component. The classic value $\Pio$, which is $0.69$ for $s=2$), is approached near the bottom left corner.

The conclusion is that $\Pi$ is generally higher in X-rays than at the lower frequencies (e.g. in the radio band). Such trend can be easily  understood if we note that: 1) the absolute value of the local spectral index around the cut-off is higher than in radio; 2) $\Pi$ is higher for larger values of $s$.
The ``local spectral index'', a concept used also in the following, is evaluated on a narrow spectral range, by fitting the local spectrum with a power law.

The increase of $\Pi$ with $s$ may approximately be verified, for the simplified case of a completely ordered MF, by using Eq.~\ref{eqs:Pio} with the ``local'' $s$ value corresponding to the local spectral index $\al$ (defined as $d\ln I/d\ln \nu$), namely $s=1-2\al$.
The exact value of the polarization degree is $\Pio$ times the ratio between the quantities $X_G$ and $X_F$. These two quantities are defined in Eqs.~\ref{eq:XFdef} and \ref{eq:XGdef}, and their ratio is also displayed in the lower panel of Fig.~\ref{fig:figintgFandGovFglob}.
This of course still in the case of a homogeneous MF, corresponding to the left edge of Figs.~\ref{fig:figPDs2bt1iso} and \ref{fig:figPDs2bt2iso}.

For the numerical evaluation of the integrals in Eqs.~\ref{eqs:aveI} and \ref{eqs:aveQ} we have used the approximations described in Appendix~\ref{app:fits2intgFG} (Eqs.~\ref{eq:XFappr} and \ref{eq:XGappr}, complemented by the coefficients in Tables~\ref{tab:XFpar} and \ref{tab:XGovFpar}).
By comparing them (for all the cases with $s=2$ and $\bt=1$), with a direct interpolation of numerically evaluated cases, we have found that the estimates of $\Pi$ obtained with the two methods coincide within a tolerance of 0.05\%, for all frequencies smaller than $100\,\omcut$; while a somehow worse accuracy appears at frequencies larger than $100\,\omcut$ (a range of much less importance, due to the rapid cutoff in the emission).

%%%%%%%%%%%%%%%%%%%%%%%%%%%%%%%%%%%%%%%%%%%%%%%%%
\subsection{The case of a purely anisotropic random field}
\label{sec:calcaniso}
%%%%%%%%%%%%%%%%%%%%%%%%%%%%%%%%%%%%%%%%%%%%%%%%%
\begin{figure} 
\centering
\includegraphics[width=.47\textwidth]{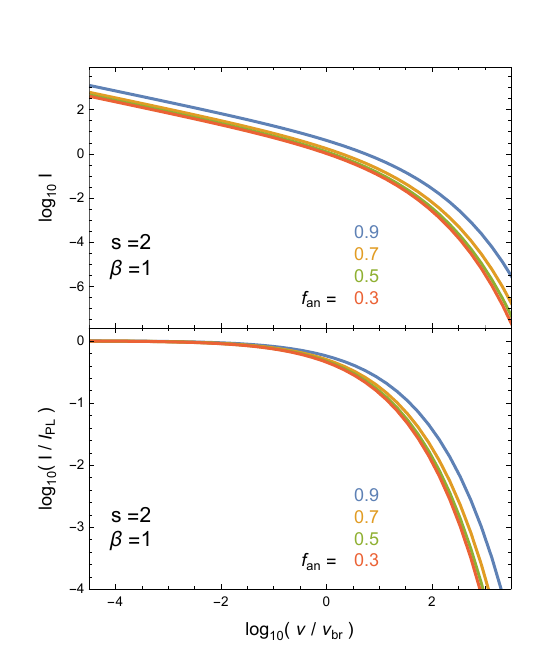}
        \caption{Spectra for ${\cal I}$, similarly to what plotted in Fig.~\ref{fig:curvesIs2bt1iso}, but now for the anisotropic case, with $\Bbar=0$. The various curves refer to different values of $\fanis$, ranging from 0.3 to 0.9 in steps of 0.2. The intensities increase with increasing $\fanis$; anyway, the differences between different models are lower than in the cases shown in Fig.~\ref{fig:curvesIs2bt1iso}, for the isotropic case. Here no black dashed curve is shown, because now the spectrum for a homogeneous MF cannot be a limit case.
        The case with $\fanis=0.0$ (isotropic random MF) is indistinguishable from the case $\fanis=0.3$.}
\label{fig:curvesIs2bt1aniso}	
\end{figure}
%%%%%%%%%%%%%%%%%%%%%%%%%%%%%%%%%%%%%%%%%%%%%%%%%
\begin{figure} 
\centering
	\includegraphics[width=.47\textwidth]{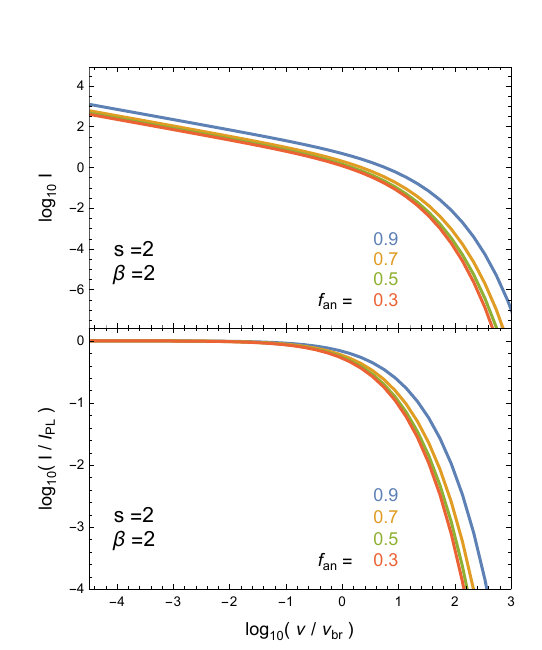}
        \caption{Same as Fig.~\ref{fig:curvesIs2bt1aniso}, but now with $\bt=2$.}
\label{fig:curvesIs2bt2aniso}
\end{figure}
%%%%%%%%%%%%%%%%%%%%%%%%%%%%%%%%%%%%%%%%%%%%%%%%%
\begin{figure} 
\centering
    \includegraphics[width=.47\textwidth]{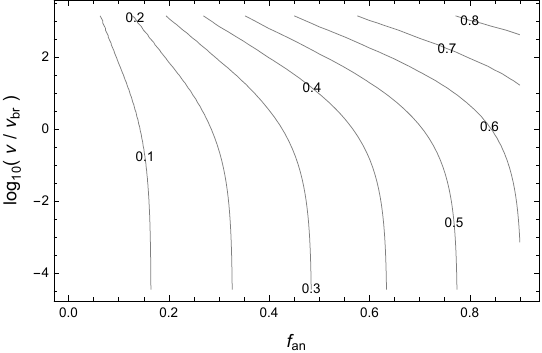}
        \caption{Iso-levels representation of the $\Pi$ dependence on the anisotropy of the random MF and on the frequency, for cases with $s=2$, $\bt=1$, and different levels of random MF anisotropy and frequency. Here only the positive values of $\fanis$ are shown, but the pattern is symmetric with respect to $\fanis=0$; changing the sign of $\fanis$ is equivalent to exchange $x$ and $y$.
        The case $\fanis=0$ is equivalent, for the results in the previous section, to the asymptotic case $\sg\gg\Bbar$.}
\label{fig:figPDs2bt1aniso}	
\end{figure}
%%%%%%%%%%%%%%%%%%%%%%%%%%%%%%%%%%%%%%%%%%%%%%%%%
\begin{figure} 
\centering
	\includegraphics[width=.47\textwidth]{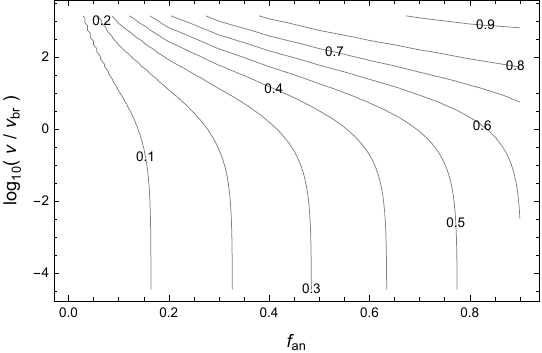}
        \caption{Same as Fig.~\ref{fig:figPDs2bt1aniso}, but now for $\bt=2$.
        }
\label{fig:figPDs2bt2aniso}	
\end{figure}
%%%%%%%%%%%%%%%%%%%%%%%%%%%%%%%%%%%%%%%%%%%%%%%%%
Another class of cases, for which in Paper~I we have found analytical solutions, consists of those with a negligible ordered MF ($\Bbar=0$) and an anisotropic random MF with symmetry axes along $x$ and $y$ (with standard deviations respectively equal to $\sg_x$ and $\sg_y$).
Following Paper~I, let us define:
%%%
\begin{equation} \sgeff^2=\frac{2\sg_x^2\sg_y^2}{\sg_x^2+\sg_y^2},
    \qquad
    \fanis=\frac{\sg_y^2-\sg_x^2}{\sg_x^2+\sg_y^2}.
\label{eq:fanis}    
\end{equation}
%%%
The defined quantity $\fanis$ ranges from $-1$ ($\sg_x\gg\sg_y$) to $+1$ ($\sg_y\gg\sg_x$). Alternatively, we may write:
%%%
\begin{equation}
    \frac{\sg_y}{\sg_x}=\sqrt{\frac{1+\fanis}{1-\fanis}}.
    \label{eq:sgyovsgxf}
\end{equation}
%%%
In a similar way to the previous section, Figs.~\ref{fig:curvesIs2bt1aniso} and \ref{fig:curvesIs2bt2aniso} show the total intensity profiles.

The most interesting behaviour is for $\Pi$, and it is shown on Figs.~\ref{fig:figPDs2bt1aniso} and \ref{fig:figPDs2bt2aniso}, for $s=2$ and, respectively, $\bt=1$ and $\bt=2$. Also in this case $\Pi$ increases with frequency around the spectral cutoff.

%%%%%%%%%%%%%%%%%%%%%%%%%%%%%%%%%%%%%%%%%%%%%%%%%
\begin{figure} 
\centering
    \includegraphics[width=.47\textwidth]{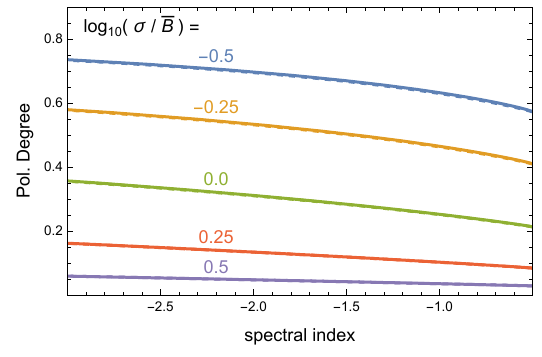}
        \caption{Dependence of $\Pi$ on the local spectral index $\al$, for $s=2$ and various levels of the isotropic random fields.
        The choice of the levels is the same as in Fig.~\ref{fig:curvesIs2bt1iso}, namely it starts with $\log_{10}\sg/\Bbar=-0.5$, and then increases in steps of $0.25$.
        $\Pi$ decreases with increasing $\sg/\Bbar$, and for values of $\log_{10}\sg/\Bbar>0.5$, the curves cannot be longer distinguished from zero in the plot. The dashed lines refer to cases with $\bt=1$, while the solid lines to $\bt=2$: the corresponding curves are almost superimposed, which means that these correlations are very insensitive to the choice of $\bt$.
        }
\label{fig:PDofSpIndxbt1andbt2iso}	
\end{figure}
%%%%%%%%%%%%%%%%%%%%%%%%%%%%%%%%%%%%%%%%%%%%%%%%%
\begin{figure} 
\centering
    \includegraphics[width=.47\textwidth]{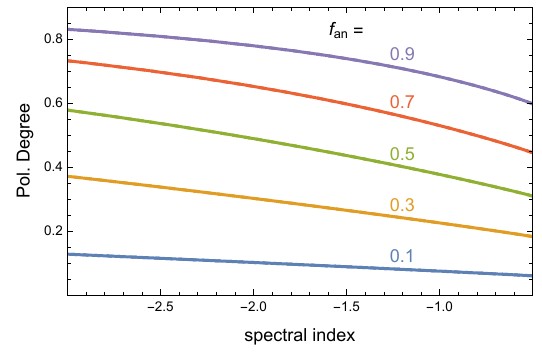}
        \caption{Same as Fig.~\ref{fig:PDofSpIndxbt1andbt2iso}, for cases of anisotropic random MF, and $\Bbar=0$. Here the choices of $\fanis$ correspond to those in Fig.~\ref{fig:curvesIs2bt1aniso}. Also in this case, the dashed lines ($\bt=1$) and the solid lines ($\bt=2$) are almost superimposed.}
\label{fig:PDofSpIndxbt1andbt2aniso}	
\end{figure}
%%%%%%%%%%%%%%%%%%%%%%%%%%%%%%%%%%%%%%%%%%%%%%%%%
\section{Observable quantities}
\label{sec:observable}

In the previous section we have shown how total emission and polarization change with frequency, there scaled with $\nubr$.
We should notice however  that, in order to derive $\nubr$ from observations,  measurements should be available over a wide spectral range (for SNRs, typically from the radio to the X-rays); in addition a fitting strategy is required, which means that the result is model-dependent (in our case, it would depend on which value for $\bt$ is assumed to start); the final best-fit values will also depend on which spectral ranges are used; last but not least, typically (at least in SNRs) the region emitting in radio is thicker than that in X-rays, which means that the emission at lower frequencies is overestimated, and as a consequence $\nubr$ is underestimated: this problem is rather general, but it will be better elucidated in Sect.~\ref{sec:youngSNRs} with reference to the case of SN~1006.

A standard model for these spectra is SRCUT \citep{1999ApJ...525..368R}, also implemented in the XSPEC package: essentially, it assumes synchrotron emission by a power-law distribution of electrons with an exponential cutoff (namely $\bt=1$, in our notation).

While one could envisage a generalization of that model to allow different values of $\bt$ as well as different levels of the random MF component,  in this section we will discuss some relations that depend only on directly observed quantities, namely: 1. the ``local spectral index'', namely that measured using only data in a narrow spectral range; 2. the polarization level in the same spectral range; 3. the spectral curvature, namely a variation with frequency of the local spectral index. Of course, interesting data are obtained only when considering sufficiently high frequencies to measure the effects of the cutoff on the particle distribution.
For SNRs, this typically happens in the X-rays spectral range, so that in the following we will explicitly refer to it.

\subsection{Polarization degree}

In Sect.~\ref{sec:calciso} we have already mentioned an increase of $\Pi$ close to the spectral cutoff and beyond. Here we will discuss in more detail the dependence of $\Pi$ on the local spectral index. 
Figure~\ref{fig:PDofSpIndxbt1andbt2iso} shows its behaviour in the case of an isotropic random MF component, for several levels of the random MF magnitude. The cases with $\bt=1$ are shown as dashed curves, while those with $\bt=2$ as solid ones. The two families of curves are almost superimposed, which means that they are very poorly sensitive to the value of $\bt$: this then represents a good measurement of the level of the random MF.
The same result comes out for the anisotropic case (and $\Bbar=0$): as shown in Fig.~\ref{fig:PDofSpIndxbt1andbt2aniso}, the two families of curves depend on the value of $\fanis$, but very poorly on the value of $\bt$, in the considered range. The regions with more extreme local spectral indices are well beyond the cutoff, so that they would be very faint and hard to be detected.

%%%%%%%%%%%%%%%%%%%%%%%%%%%%%%%%%%%%%%%%%%%%%%%%%
\begin{figure} 
\centering
    \includegraphics[width=.47\textwidth]{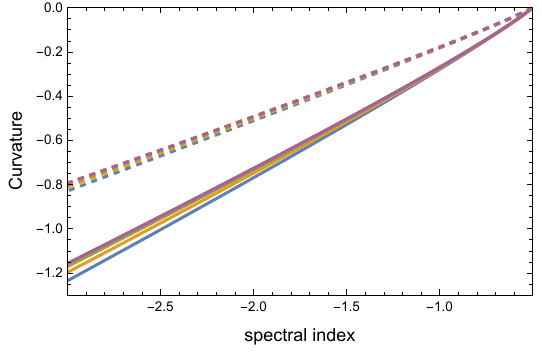}
        \caption{Dependence of the curvature on the local spectral index $\al$, for $s=2$ and various levels of the isotropic random fields (same choice of levels as in Fig.~\ref{fig:curvesIs2bt1iso}). The dashed lines refer to cases with $\bt=1$, while the solid lines to $\bt=2$. Note that, in this case, the trends are rather insensitive to the level of the random MF, while they are mostly sensitive to the value of $\bt$.}
\label{fig:CURVofSpIndxbt1andbt2iso}	
\end{figure}
%%%%%%%%%%%%%%%%%%%%%%%%%%%%%%%%%%%%%%%%%%%%%%%%%
\begin{figure} 
\centering
    \includegraphics[width=.47\textwidth]{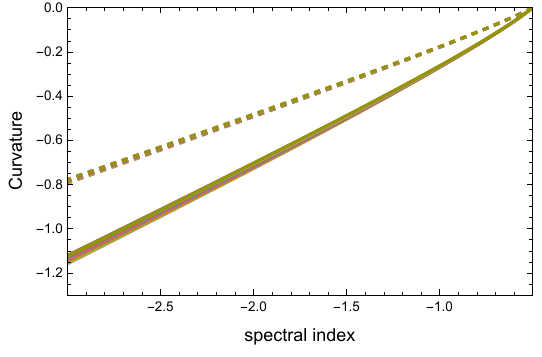}
        \caption{Same as Fig.~\ref{fig:CURVofSpIndxbt1andbt2iso}, for cases of anisotropic random MF (plus $\Bbar=0$). The choice of colors is the same as in Fig.~\ref{fig:curvesIs2bt1aniso}. Also in this case the dependence on the level of anisotropy is low, while that to the value of $\bt$ is higher.}
\label{fig:CURVofSpIndxbt1andbt2aniso}	
\end{figure}
%%%%%%%%%%%%%%%%%%%%%%%%%%%%%%%%%%%%%%%%%%%%%%%%%
\subsection{Spectral curvature}
\label{sec:curvature}

A further observable could be considered, even if more difficult to measure, namely the curvature of the spectrum in the X-ray region, defined here as the derivative of the local spectral index with respect to the logarithm of the frequency $d \al/d\ln\nu$.
It requires a wide X-ray spectral range to be measured, while for the main imaging telescopes the width of the X-ray spectral window does not extend beyond $\simeq10\U{keV}$.
Only recently the required conditions have been accomplished by combining for instance XMM-Newton and NuSTAR data \citep[see e.g.][]{2018ApJ...864...85L}.

Figures~\ref{fig:CURVofSpIndxbt1andbt2iso} and \ref{fig:CURVofSpIndxbt1andbt2aniso} show that the spectral curvature (i.e. how much the \textit{shape} of the synchrotron spectrum deviates from a pure power law), as a function of the local spectral index, is sensitive to the type of cutoff (namely to the parameter $\bt$), while it is almost insensitive to the level of the random field component, both in the isotropic and the anisotropic case (with $\Bbar=0$).
Therefore, the measurement of the spectral curvature could be very effective for investigating the physical nature of the spectral cutoff.
The only problem could be in principle that the emitting region could not be exactly the same at all X-ray frequencies: a caveat similar to that about the comparison between radio and X-ray emission, but now much less relevant because the spectral lever is hardly wider than just one decade.

%%%%%%%%%%%%%%%%%%%%%%%%%%%%%%%%%%%%%%%%%%%%%%%%%
\begin{figure} 
\centering
    \includegraphics[width=.47\textwidth]{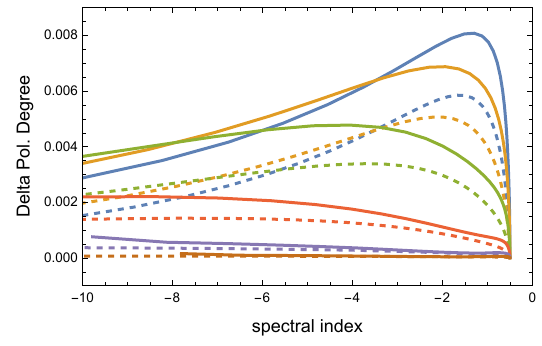}
        \caption{Difference between the numerically evaluated $\Pi$ and the analytic approximation for a power law distribution, for all of our models with an isotropic random MF. Note that the difference is always lower than 0.01.}
\label{fig:RESIDPDofSpIndxbt1andbt2iso}	
\end{figure}
%%%%%%%%%%%%%%%%%%%%%%%%%%%%%%%%%%%%%%%%%%%%%%%%%
\begin{figure} 
\centering
    \includegraphics[width=.47\textwidth]{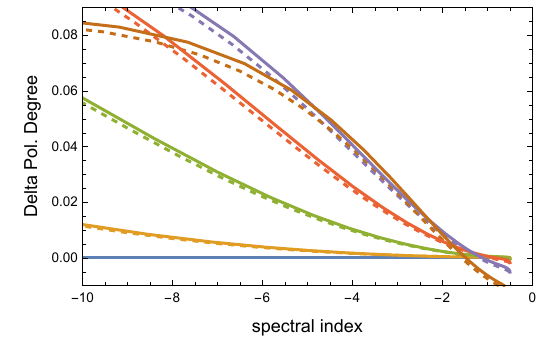}
        \caption{Same as Fig.~\ref{fig:RESIDPDofSpIndxbt1andbt2iso}, but for the case of an anisotropic random MF (plus $\Bbar=0$). Note that also in this case the difference is almost always lower than 0.01.}
\label{fig:RESIDPDofSpIndxbt1andbt2aniso}	
\end{figure}
%%%%%%%%%%%%%%%%%%%%%%%%%%%%%%%%%%%%%%%%%%%%%%%%%
\subsection{Accuracy of the analytic formulae}
\label{sec:accuracyapprox}

One could wonder how well a real case with curved spectrum in X-rays can be approximated by the analytic solutions for a purely power-law distribution (Paper~I)), both in the isotropic case and in the anisotropic one (and $\Bbar=0$). Figure~\ref{fig:RESIDPDofSpIndxbt1andbt2iso} shows, for our models with an isotropic random MF, the difference between the numerically evaluated $\Pi$ and the analytic formulae from our Paper~I, when applied to the local spectral index, after fitting the local spectrum with a power law.
Even if the spectrum is bent, because of its proximity to the cutoff (as it is usually the case for SNRs in the X-ray spectral range), the figure proves that our analytic formulae are generally very good approximations of the real cases.
The numerically determined values of $\Pi$ are always larger than the analytic approximations, but with a discrepancy always smaller than 0.01. 
A similar conclusion can be obtained also for the case of an anisotropic random MF, as shown in Fig.~\ref{fig:RESIDPDofSpIndxbt1andbt2aniso}.

%%%%%%%%%%%%%%%%%%%%%%%%%%%%%%%%%%%%%%%%%%%%%%%%%
\begin{figure} 
\centering
    \includegraphics[width=.47\textwidth]{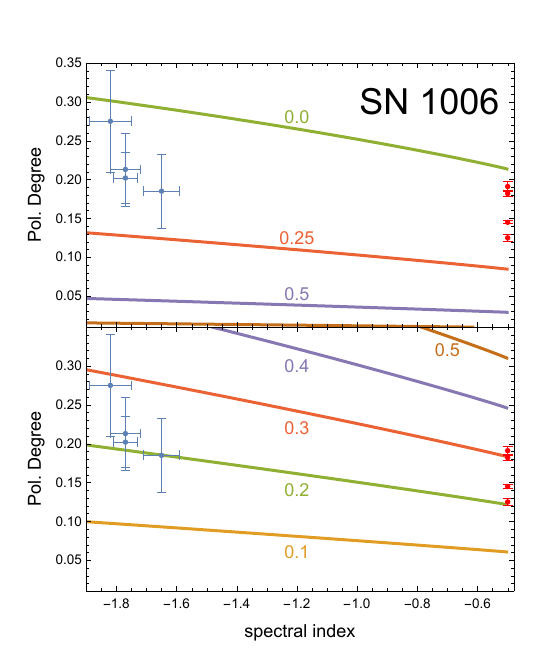}
        \caption{Comparison of the radio and X-ray polarization data on SN~1006 \citet[data as from][]{2023ApJ...957...55Z}. The upper panel shows them superimposed to models of a homogeneous MF composed with an isotropic random component (the labels indicate $\log_{10}\sg/\Bbar$). The lower panel compares instead them with models of an anisotropic random MF (the labels now indicate $\fanis$).}
\label{fig:comparewithIXPE}	
\end{figure}
%%%%%%%%%%%%%%%%%%%%%%%%%%%%%%%%%%%%%%%%%%%%%%%%%
\section{Application to some young SNRs}
\label{sec:youngSNRs}

The young supernova remnant SN~1006, with its wide angular size and rather ordered morphology, represents a optimal source for a spatially resolved analysis of synchrotron radiation, both in the radio and in the X-rays (also this one dominated by non-thermal emission).

To get a reliable estimate of the cutoff frequency ($\nubr$ in our notation), one would need to measure the emission from a fixed volume element, with homogeneous physical conditions.
Instead, by a comparative analysis of the radio and X-ray patterns for SN~1006 \cite[see e.g.][]{2004A&A...425..121R,2009MNRAS.399..157P}, one can infer that the X-ray limb is much thinner than that in radio.
This difference is likely due to the different synchrotron lifetimes of the emitting electrons, in the downstream, and it is a common characteristics of young SNRs.

This means that (in addition to other possible changes of conditions) the emitting domain along a given LoS is longer at lower frequencies: close to the projected edge of the SNRs the effect of this bias could be marginal, but it is more consistent for a LoS not so close to the SNR edge.
The result is to underestimate the X-ray emissivity with respect to the radio one, and consequently to underestimate the cutoff frequency.
This effect easily justifies the maps of cutoff frequency presented in some papers, among which \citet{2010ApJ...723..383K} and \citet{2018ApJ...864...85L}, which show a strong radial gradient for this quantity.
A similar effect takes place also in the presence of MF inhomogeneities inside the SNR, even though under usual conditions the main effect is expected to derive from the energy dependence of the synchrotron radiative time.

The least biased measurements of $\nubr$ are those taken very close to the SNR projected edge; even though a slight wrinkling at the edge would be enough to distort those measurements as well.
This case also justifies why, in our previous modelling, we have searched for relations between actually observed quantities, and that do not depend on the actual value of $\nubr$, a quantity hard to evaluate in an unbiased and model independent way. 

Using the polarization measurements in \citet{2023ApJ...957...55Z}  (their Table~1), in Fig.~\ref{fig:comparewithIXPE} we show that $\Pi$ in radio and that in the X-rays agree with each other, 
which could mean that, in spite of the different extensions of the emitting region in these two spectral ranges, the data are consistent with the properties of the random MF component being the same.
In the upper panel the data are superimposed to models of a homogeneous MF + an isotropic random component: the various data are compatible with $\sg/\Bbar\simeq1.0\div1.4$ (note that in that paper the symbol $\sg_B$ is used for our $\sg$).
Of course, the values of $\sg$ for an isotropic random field are upper limits, since they are estimated in the limit case that $\Bbar$ keeps constant along the whole LoS.
As already mentioned by \citet{2023ApJ...957...55Z}, these values are lower than the magnetic amplification expected theoretically, and in some cases derived from the sharp profiles of the nonthermal X-ray filaments.

In the lower panel, instead, the data are superimposed to models of an anisotropic random MF, at a level much higher than any ordered background MF: in this case the data are compatible with $\fanis\simeq0.2\div0.3$, with the width of the probability function larger in the radial direction.
In order to have a $\fanis$ with positive sign, consistently with Eq.~\ref{eq:fanis}, we assume in the present section the $y$ direction to be parallel to the shock velocity (from which $\sg_y\Rightarrow\sgpar$), while the $x$ direction perpendicular to it (from which $\sg_x\Rightarrow\sgperp$).
In other terms, we can measure the ratio $\sgpar/\sgperp$ ($>1$) in the downstream, by using $\fanis$ as derived from actual measurements of $\Pi$ and $\alpha$.

While with the present-day polarization data there is no reason to prefer one model to the other, one may note that in the first case the estimated level of the random MF would not comply with a high MF amplification in the upstream, usually taken as a necessary condition for having a high acceleration efficiency, as in the case for SN~1006.
On the other hand, our second model is compatible with the idea that the shock is an efficient cosmic-ray accelerator, and for this reason we think it has to be preferred.

A random MF preferentially oriented along the radial direction is more likely understood as the effect of motions, driven by some hydrodynamic instability. A well known mechanism is Rayleigh-Taylor instability, which occurs when the density is gradient opposite to the pressure gradient. The role of this instability on the development of radially oriented MFs in SNRs has been first discussed in detail by \citet{1996ApJ...472..245J}. On the other hand the Rayleigh-Taylor instability resulted to be mostly effective only near the contact discontinuity, while radially oriented MFs are observed also in the region ahead of the contact discontinuity, up to close to the forward shock. The Richtmyer-Meshkov instability then resulted to be more promising to explain the observed phenomenology. This instability originates from corrugations of the shock surface, related to inhomogeneities in the ambient medium. Its role in generating radially oriented MFs has been investigated by \citet{2013ApJ...772L..20I}. A detailed analysis of these two instabilities is beyond the scope of the present work; while for more detail one may refer, for instance, to the review by \citet{2021PhyD..42332838Z}.

More generally, a disordered magnetic field could be generated in the upstream as well as in the downstream.
In the case some random field (with dispersions $\sgupar$ and $\sguperp$) is already present in the upstream, after the compression at the shock the perpendicular spread would change to $\kappa\,\sguperp$, where $\kappa$ is the shock compression factor.
Then, given an observed random MF downstream with dispersions $\sgpar$ and $\sgperp$, the random MF generated in the downstream should be with:
%%%
\begin{eqnarray}
    \sggpar&=&\sqrt{\sgpar^2-\sgupar^2};\\
    \sggperp&=&\sqrt{\sgperp^2-{\kappa^2\sguperp^2}}.
\end{eqnarray}
%%%
It may be seen that, for a negligible random field in the upstream, the spread ratio $\sggpar/\sggperp$ for the generated random MF is equal to the measured $\sgpar/\sgperp$. Instead, for an isotropic random field in the upstream, there should be $\sggpar/\sggperp>\sgpar/\sgperp$. On the opposite, without MF generation in the downstream, in order to have a measured $\sgpar/\sgperp>1$ in the downstream, a highly anisotropic random field would be required in the pre-shock region, with $\sgupar/\sguperp>\kappa$.
The fact that the resulting orientation of this anisotropy is radial in SN~1006, contrary to what expected in the case of a mere shock compression of a pre-existing (isotropic) random MF, then suggests that the mechanism driving a radial orientation of this MF downstream must be very efficient.

In a similar way, one can extract some information from X-ray polarization measurements of Tycho.
\citet{2023ApJ...945...52F}, on the base of IXPE data, obtained measures of the X-ray polarization degree $\Pi$, and the local spectral index $\al$, in some selected regions. Among these: $\Pi=11.9\pm2.2\%$ and $\al=-1.82\pm0.02$ in the rim region; $\Pi=23.4\pm4.2\%$ and $\al=-1.90\pm0.04$ in a west region (dubbed (f) in that paper).
Using the formulae from our Paper I, \citet{2023ApJ...945...52F} have derived, respectively, $\dl B/\Bbar=3.3\pm0.4$ and $\dl B/\Bbar=2.1\pm0.3$. Note that in that paper the quantity $\dl B=\sqrt{3}\,\sg$ is used\footnote{The authors consider $\dl B$ as the dispersion of the module of the random MF vector, in the 3-D space, while our $\sg$ corresponds to that of its 1-D component. Also, the quantity $B$ in that paper corresponds to our $\Bbar$, namely it is the 2-D projection of the 3-D ordered MF component (let us call it $B\rs{o}$). If the value $\dl B/B\rs{o}$ refers to the ratio of the magnitudes of the 3-D vectors, then the general way to convert our $\sg/\Bbar$ to this ratio is $\dl B/B\rs{o}=\sqrt{3}\sin\phi\cdot \sg/\Bbar$ where $\phi$ is the angle between the LoS and $B\rs{o}$. 
If one considers a radial ordered MF and regions close to the edge of the SNR projection then $\sin\phi=1$.}, while in our notation those measurements correspond respectively to $\sg/\Bbar=1.9\pm0.2$ and to $\sg/\Bbar=1.2\pm0.2$. Note that also in this case a moderate MF amplification is estimated, contrary to the large amplification theoretically required.
On the other hand, a dominant anisotropic random field could reproduce the data, provided that the anisotropy factor is $\fanis=0.12\pm0.02$ in the first case, and $\fanis=0.24\pm0.04$ in the second case.
According to Eq.~\ref{eq:sgyovsgxf}, these values correspond respectively to $\sgpar/\sgperp=1.13\pm0.03$ and $1.28\pm0.06$.
Unfortunately, differently from \citet{2023ApJ...957...55Z}, in this paper there are no comparative results on $\Pi$ in radio for the same regions, so that one cannot display for Tycho a figure similar to Fig.~\ref{fig:comparewithIXPE}.

Finally, \citet{2022ApJ...938...40V} reported the results of IXPE observations of Cas~A. Unfortunately, given the small angular size of this SNR, it cannot be resolved as well as in the other two cases. In order to optimise the spatial information, the authors have imposed a circular symmetry for the polarization vectors, and have corrected for the thermal X-ray emission. Even so, the X-ray polarization degree is in the range $\sim2$--$5\%$, namely lower than the $\sim5\%$ in the radio band, a result apparently impossible, since the spectral index in X-rays is steeper than that in radio. If this result of the data analysis will be confirmed, the most reasonable explanation could be that the radio emission comes from a layer substantially thicker than the X-rays, so that the radio and X-ray emitting electrons interact with random fields having either different $\sigma/\Bbar$ or different $\fanis$. Apart from this general consideration, however, the quality of these data is not sufficient to perform any accurate model fitting.

%%%%%%%%%%%%%%%%%%%%%%%%%%%%%%%%%%%%%%%%%%%%%%%%%
\begin{figure} 
\centering
    \includegraphics[width=.47\textwidth]{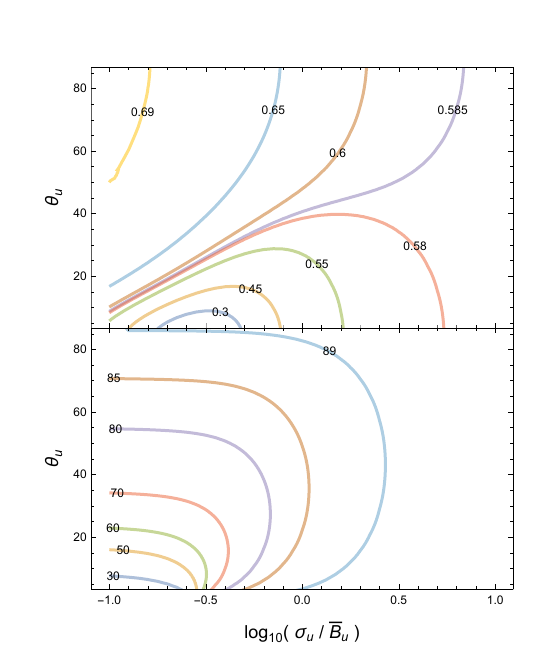}
        \caption{Map of $\Pi$ (upper panel) and of the (magnetic) polarization direction $\tht\rs{s}$ (in degrees, from the direction of the shock velocity), for a wide choice of upstream MF conditions. Here the particle distribution has a power-law index $s=2$, corresponding to a spectral index $\al=-0.5$.}
\label{fig:figPLxPDandTHTs2}	
\end{figure}
%%%%%%%%%%%%%%%%%%%%%%%%%%%%%%%%%%%%%%%%%%%%%%%%%
\begin{figure} 
\centering
    \includegraphics[width=.47\textwidth]{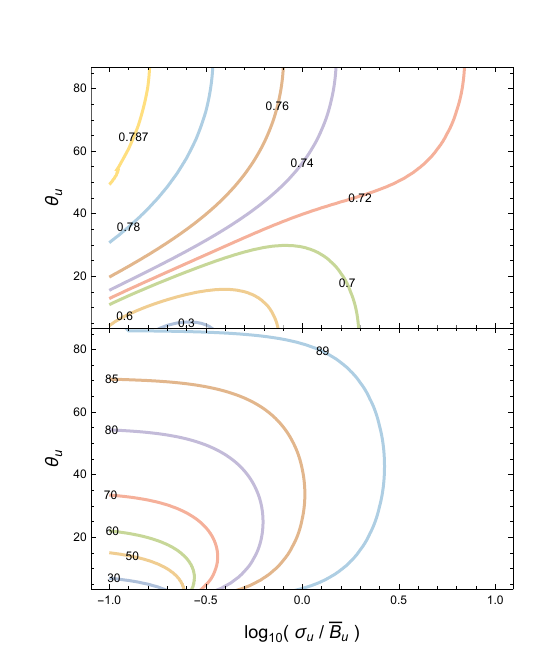}
        \caption{Same as Fig.~\ref{fig:figPLxPDandTHTs2}, with a particle power-law index $s=4$, corresponding to a spectral index $\al=-1.5$.}
\label{fig:figPLxPDandTHTs4}	
\end{figure}
%%%%%%%%%%%%%%%%%%%%%%%%%%%%%%%%%%%%%%%%%%%%%%%%%
\begin{figure} 
\centering
    \includegraphics[width=.47\textwidth]{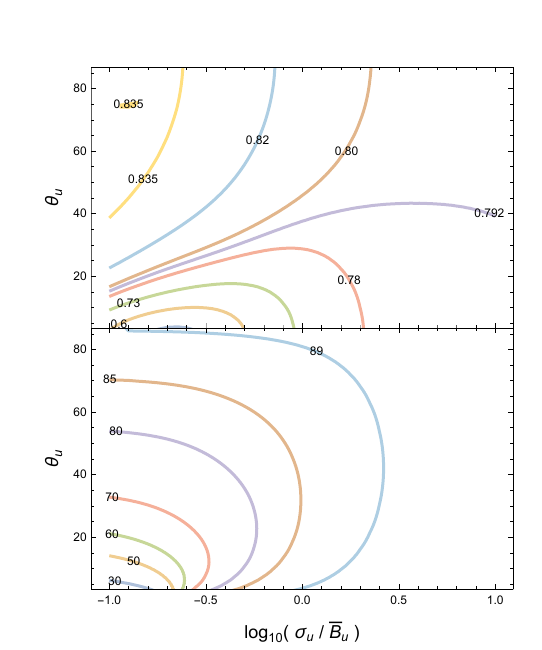}
        \caption{Same as Fig.~\ref{fig:figPLxPDandTHTs2}, with a particle power-law index $s=6$, corresponding to a spectral index $\al=-2.5$.}
\label{fig:figPLxPDandTHTs6}	
\end{figure}
%%%%%%%%%%%%%%%%%%%%%%%%%%%%%%%%%%%%%%%%%%%%%%%%%
\section{Homogeneous + anisotropic random field}
\label{sec:homo+anis}

The result given in Sect.~\ref{sec:accuracyapprox} is on one side comforting, because is proves that the original analytic formulae (from our Paper~I) allow one to reach a good accuracy with a minimum effort; on the other hand, it may appear rather frustrating if all the numerical machinery developed in the present work has served just to test the level of accuracy of our previous analytic functions, when applied to the X-ray polarization, by using the local spectral index.
In fact this is not the case, because the numerical method presented here allows one also to tackle a much wider variety of problems that could not be investigated analytically.

As an example, let us model here the case of a non negligible homogeneous MF combined with an anisotropic random MF.
In particular, we will consider the case of a strong shock, with compression factor $\kappa=4$, moving through an ambient medium characterised upstream by a homogeneous MF component ($\Bbaru$),  combined with an isotropic random component ($\sgu$). We consider here the case when the only effect of the shock is related to the compression of matter and MF. This should be a rather common case for not too young SNRs.

After the shock passage, in the downstream the field component along the shock velocity stays unchanged, while those orthogonal to it would be enhanced by a factor $\kappa$, both for the ordered component ($\Bbardpar=\Bbarupar$, $\Bbardperp=\kappa\,\Bbaruperp$) and the random one ($\sgdpar=\sgupar$, $\sgdperp=\kappa\,\sguperp$).
The case with a negligible homogeneous component has been already analyzed in the Sect.~2.3 of Paper I. 
Even in the simplified case of a power-law distribution, with respect to Paper I here we have to add 2 further parameters describing the MF conditions upstream: the ratio between random and ordered MF magnitudes ($\sgu/\Bbaru$), and the direction of the ordered MF with respect to the direction of the shock velocity ($\thtu$), both in the upstream.

Here we restrict our analysis to pure power-law particle energy distributions, but with a similar effort one could consider also the case with a spectral cutoff. Anyway, as already shown in Sect.~\ref{sec:accuracyapprox} for similar cases, results valid for pure power-law distributions could be accurate enough also in the more general case, simply using the ``local spectral index''. 
Figures~\ref{fig:figPLxPDandTHTs2}, \ref{fig:figPLxPDandTHTs4}, and \ref{fig:figPLxPDandTHTs6} display the results for power-law particle energy distributions with indices $s=2$, $s=4$, and $s=6$ respectively: these cases corresponding to spectral indices $\al=-0.5$, $-1.5$, and $-2.5$.
Each figure shows both the level of $\Pi$ (upper panel) and the direction of the (magnetic) polarization with respect to the normal to the shock surface.
 
As one can notice, the qualitative trends in these three figures are similar.  In particular, the lower panels show very similar behaviours, with differences noticeable only for small values of $\thtd$. These trends can be interpreted as follows: in the case a very low level of the random MF upstream ($\sgu\ll\Bbaru$), $\Pi$ approaches the limit value for a homogeneous MF (Eq.~\ref{eqs:Pio}), while its direction is always larger that $\thtu$, being the MF direction downstream $\thtd=\arctan(\kappa\,\tan(\thtu))$; instead, in the case $\sgu\gg\Bbaru$, $\thtd$ is always very close to $90^\circ$, while $\Pi$ approaches the analytic values given by Eq.~42 in Paper I; in the intermediate cases one may see that, at fixed $\thtu$, $\thtd$ increases monotonically with $\sgu/\Bbaru$; the value $\Pi$, instead, is monotonically decreasing only for large enough values of $\thtu$; for smaller values of $\thtu$, instead, the behavior of $\Pi$ with $\sgu/\Bbaru$ experiences a minimum, essentially when the ordered and the random MF components are of comparable strength, but one is predominantly parallel to the shock velocity, while the other is perpendicular to it, so that their different polarizations tend to eliminate each other.

%%%%%%%%%%%%%%%%%%%%%%%%%%%%%%%%%%%%%%%%%%%%%%%%%
\section{Conclusions}
\label{sec:conclusions}

In this paper, which is intended to be a continuation of our Paper I, after having outlined the basics of our general theory, we have described a numerical implementation that allows one to calculate in an efficient way the spectral behaviour of the synchrotron emission and of its polarization degree, for a wide variety of cases in which an ordered MF component is complemented with a random one.

Our aim is to extend the application of the theory to the X-ray polarization.
We have first performed an analysis of the two cases already treated in Paper I, namely, an isotropic random MF plus an ordered MF and the anisotropic random MF with vanishing ordered one, but now for a particle energy distribution with an exponential / super-exponential cutoff. While the ``true'' cutoff frequency is hard to measure, we have discussed the relationships between directly observable quantities, like the local spectral index (namely that computed in a limited spectral band), the local spectral curvature, and the polarization degree.
We have also shown that, with a reasonable level of approximation, our former analytic formulae computed for a pure power-law case could be used also in a more general case, by applying them to the local spectral index, also in the presence of a cutoff.
We have then applied some of our findings to the X-ray polarization results, based on IXPE data, for the young remnants of supernovae SN~1006, Tycho and Cas A.
Finally, we have treated the case of a mixture of homogeneous and anisotropic random MF, specialised to the case in which a strong shock compresses an ambient field with also a random component: this case should be of particular interest for the radio emission from mid-age SNRs.
As already discussed in Sect.~\ref{sec:method}, the present treatment assumes a large number of MF realizations along a LoS. Moreover, in the presence of MF gradients in the SNR interior as a result of the MHD evolution of plasma, like in \citet{2017MNRAS.470.1156P}, this condition must apply also for the length scale of such variations.

The extension to the X-ray band has made polarimetry an even more important diagnostic tool for studying synchrotron emission from astrophysical sources.

%%%%%%%%%%%%%%%%%%%%%%%%%%%%%%%%%%%%%%%%%%%%%%%%%
\begin{acknowledgements}
This work has been partially funded by the European Union - Next Generation EU, through PRIN-MUR 2022TJW4EJ. RB also acknowledges support from the Italian National Institute for Astrophysics with PRIN-INAF 2019 and MiniGrants PWNnumpol and HYPNOTIC87A.
OP acknowledges the OAPa grant number D.D.75/2022 funded by Direzione Scientifica of Istituto Nazionale di Astrofisica, Italy. This project has received funding through the MSCA4Ukraine project, which is funded by the European Union. Views and opinions expressed are however those of the authors only and do not necessarily reflect those of the European Union. Neither the European Union nor the MSCA4Ukraine Consortium as a whole nor any individual member institutions of the MSCA4Ukraine Consortium can be held responsible for them.
The numerical and analytical calculations presented in this article, as well as the figures shown, were obtained using Wolfram Mathematica.
\end{acknowledgements}

\bibliographystyle{aa}
\bibliography{biblio}

\begin{thebibliography}{19}
\expandafter\ifx\csname natexlab\endcsname\relax\def\natexlab#1{#1}\fi

\bibitem[{{Bandiera} \& {Petruk}(2016)}]{2016MNRAS.459..178B}
{Bandiera}, R. \& {Petruk}, O. 2016, \mnras, 459, {178 (Paper I)}

\bibitem[{{Bucciantini} {et~al.}(2023){Bucciantini}, {Ferrazzoli}, {Bachetti}, {Rankin}, {Di Lalla}, {Sgr{\`o}}, {Omodei}, {Kitaguchi}, {Mizuno}, {Gunji}, {Watanabe}, {Baldini}, {Slane}, {Weisskopf}, {Romani}, {Possenti}, {Marshall}, {Silvestri}, {Pacciani}, {Negro}, {Muleri}, {de O{\~n}a Wilhelmi}, {Xie}, {Heyl}, {Pesce-Rollins}, {Wong}, {Pilia}, {Agudo}, {Antonelli}, {Baumgartner}, \& et~al.}]{2023NatAs...7..602B}
{Bucciantini}, N., {Ferrazzoli}, R., {Bachetti}, M., {et~al.} 2023, Nature Astronomy, 7, 602

\bibitem[{{Churazov}(2023)}]{2023arXiv231204421C}
{Churazov}, E. 2023, arXiv e-prints, arXiv:2312.04421

\bibitem[{{Ferrazzoli} {et~al.}(2023){Ferrazzoli}, {Slane}, {Prokhorov}, {Zhou}, {Vink}, {Bucciantini}, {Costa}, {Di Lalla}, {Di Marco}, {Soffitta}, {Weisskopf}, {Asakura}, {Baldini}, {Heyl}, {Kaaret}, {Marin}, {Mizuno}, \& et~al.}]{2023ApJ...945...52F}
{Ferrazzoli}, R., {Slane}, P., {Prokhorov}, D., {et~al.} 2023, \apj, 945, 52

\bibitem[{{Inoue} {et~al.}(2013){Inoue}, {Shimoda}, {Ohira}, \& {Yamazaki}}]{2013ApJ...772L..20I}
{Inoue}, T., {Shimoda}, J., {Ohira}, Y., \& {Yamazaki}, R. 2013, \apjl, 772, L20

\bibitem[{{Jun} \& {Norman}(1996)}]{1996ApJ...472..245J}
{Jun}, B.-I. \& {Norman}, M.~L. 1996, \apj, 472, 245

\bibitem[{{Katsuda} {et~al.}(2010){Katsuda}, {Petre}, {Mori}, {Reynolds}, {Long}, {Winkler}, \& {Tsunemi}}]{2010ApJ...723..383K}
{Katsuda}, S., {Petre}, R., {Mori}, K., {et~al.} 2010, \apj, 723, 383

\bibitem[{{Li} {et~al.}(2018){Li}, {Ballet}, {Miceli}, {Zhou}, {Vink}, {Chen}, {Acero}, {Decourchelle}, \& {Bregman}}]{2018ApJ...864...85L}
{Li}, J.-T., {Ballet}, J., {Miceli}, M., {et~al.} 2018, \apj, 864, 85

\bibitem[{{Liu} {et~al.}(2023){Liu}, {Xie}, {Liu}, {Ng}, {Bucciantini}, {Romani}, {Weisskopf}, {Costa}, {Di Marco}, {La Monaca}, {Muleri}, {Soffitta}, {Deng}, {Meng}, \& {Liang}}]{2023ApJ...959L...2L}
{Liu}, K., {Xie}, F., {Liu}, Y.-h., {et~al.} 2023, \apjl, 959, L2

\bibitem[{{Petruk} {et~al.}(2017){Petruk}, {Bandiera}, {Beshley}, {Orlando}, \& {Miceli}}]{2017MNRAS.470.1156P}
{Petruk}, O., {Bandiera}, R., {Beshley}, V., {Orlando}, S., \& {Miceli}, M. 2017, \mnras, 470, 1156

\bibitem[{{Petruk} {et~al.}(2009){Petruk}, {Bocchino}, {Miceli}, {Dubner}, {Castelletti}, {Orlando}, {Iakubovskyi}, \& {Telezhinsky}}]{2009MNRAS.399..157P}
{Petruk}, O., {Bocchino}, F., {Miceli}, M., {et~al.} 2009, \mnras, 399, 157

\bibitem[{{Reynolds} \& {Keohane}(1999)}]{1999ApJ...525..368R}
{Reynolds}, S.~P. \& {Keohane}, J.~W. 1999, \apj, 525, 368

\bibitem[{{Rothenflug} {et~al.}(2004){Rothenflug}, {Ballet}, {Dubner}, {Giacani}, {Decourchelle}, \& {Ferrando}}]{2004A&A...425..121R}
{Rothenflug}, R., {Ballet}, J., {Dubner}, G., {et~al.} 2004, \aap, 425, 121

\bibitem[{{Rybicki} \& {Lightman}(1986)}]{1986rpa..book.....R}
{Rybicki}, G.~B. \& {Lightman}, A.~P. 1986, {Radiative Processes in Astrophysics}

\bibitem[{{Vink} {et~al.}(2022){Vink}, {Prokhorov}, {Ferrazzoli}, {Slane}, {Zhou}, {Asakura}, {Baldini}, {Bucciantini}, {Costa}, {Di Marco}, {Heyl}, {Marin}, {Mizuno}, {Ng}, {Pesce-Rollins}, {Ramsey}, {Rankin}, {Ratheesh}, {Sgr{\'o}}, {Soffitta}, {Swartz}, {Tamagawa}, {Weisskopf}, {Yang}, {Bellazzini}, {Bonino}, {Cavazzuti}, {Costamante}, {Di Lalla}, {Latronico}, {Maldera}, {Manfreda}, {Massaro}, {Mitsuishi}, {Omodei}, {Oppedisano}, {Zane}, {Agudo}, {Antonelli}, {Bachetti}, {Baumgartner}, {Bianchi}, {Bongiorno}, {Brez}, {Capitanio}, {Castellano}, {Ciprini}, {De Rosa}, {Del Monte}, {Di Gesu}, {Donnarumma}, {Doroshenko}, {Dov{\v{c}}iak}, {Ehlert}, {Enoto}, {Evangelista}, {Fabiani}, {Garcia}, {Gunji}, {Hayashida}, {Iwakiri}, {Jorstad}, {Karas}, {Kitaguchi}, {Kolodziejczak}, {Krawczynski}, {La Monaca}, {Liodakis}, {Marinucci}, {Marscher}, {Marshall}, {Matt}, {Muleri}, {O'Dell}, {Papitto}, {Pavlov}, {Peirson}, {Perri}, {Pilia}, {Possenti}, {Poutanen}, {Puccetti}, {Romani}, {Spandre}, {Tavecchio}, {Taverna},
  {Tawara}, {Tennant}, {Thomas}, {Tombesi}, {Trois}, {Tsygankov}, {Turolla}, {Wu}, \& {Xie}}]{2022ApJ...938...40V}
{Vink}, J., {Prokhorov}, D., {Ferrazzoli}, R., {et~al.} 2022, \apj, 938, 40

\bibitem[{{Weisskopf}(2022)}]{2022HEAD...1930101W}
{Weisskopf}, M. 2022, in AAS/High Energy Astrophysics Division, Vol.~54, AAS/High Energy Astrophysics Division, 301.01

\bibitem[{{Zhou} {et~al.}(2023){Zhou}, {Prokhorov}, {Ferrazzoli}, {Yang}, {Slane}, {Vink}, {Silvestri}, {Bucciantini}, {Reynoso}, {Moffett}, {Soffitta}, {Swartz}, {Kaaret}, {Baldini}, {Costa}, {Ng}, {Kim}, {Doroshenko}, {Ehlert}, {Heyl}, {Marin}, \& et~al.}]{2023ApJ...957...55Z}
{Zhou}, P., {Prokhorov}, D., {Ferrazzoli}, R., {et~al.} 2023, \apj, 957, 55

\bibitem[{{Zhou} {et~al.}(2021){Zhou}, {Williams}, {Ramaprabhu}, {Groom}, {Thornber}, {Hillier}, {Mostert}, {Rollin}, {Balachandar}, {Powell}, {Mahalov}, \& {Attal}}]{2021PhyD..42332838Z}
{Zhou}, Y., {Williams}, R. J.~R., {Ramaprabhu}, P., {et~al.} 2021, Physica D Nonlinear Phenomena, 423, 132838

\bibitem[{{Zirakashvili} \& {Aharonian}(2007)}]{2007A&A...465..695Z}
{Zirakashvili}, V.~N. \& {Aharonian}, F. 2007, \aap, 465, 695

\end{thebibliography}

%TC:ignore
\begin{appendix}

%%%% Appendix A EXT %%%%%%%%%%%%%%%%%%%%%%%%%%%%%
\section{Fits to the $F(x)$ and $G(x)$ functions}
\label{app:fits2FG}
In detailed calculations of the synchrotron emission one needs to use two special functions, $F(x)$ and $G(x)$ (with $x\geq0$), defined by Eqs.~\ref{eq:Fdef} and \ref{eq:Gdef} in the main text.
These functions, based on modified Bessel functions of the second kind, and on a related integral, are rather heavy to compute numerically.
Since the standard asymptotic approximations:
%%%
\begin{eqnarray}
    F_{x\ll1}(x)\,\,\simeq\,\,2\,G_{x\ll1}(x)&\simeq&\frac{4\upi}{\sqrt{3}\,\Gm(1/3)}\left(\frac{x}{2}\right)^{1/3};   \\
    F_{x\gg1}(x)\,\,\simeq\quad G_{x\gg1}(x)&\simeq&\left(\frac{\upi}{2}x\right)^{1/2}e^{-x}
\end{eqnarray}
%%%
are limited to respectively very small or very large values of $x$, we have devised fits that, joined together, are valid over the whole range of the argument $x$, and allow a very high accuracy ($\sim10^{-6}$ in the worst case).

For $F(x)$ and $x<1$ we have applied to the asymptotic approximation the following correction factor:
%%%
\begin{eqnarray}
   F_{\rm{corr,\,} x<1}(x)&\simeq&\!\!\!\!\!\!
    1-0.843813 x^{2/3}+0.1875 x^2-0.066244 x^{10/3}
    \nonumber\\ &&\!\!\!\!\!\!\!\!\!\!\!\!\!\!\!\!\!\!\!\!\!\!\!\!\!\!\!\!\!\!\!\!\!\!\!\!\!\!\!\!\!\!\!\!\!\!\!\!
    +0.028125
   x^4-0.00354879 x^{16/3}+0.00109863 x^6
    \\    &&\!\!\!\!\!\!\!\!\!\!\!\!\!\!\!\!\!\!\!\!\!\!\!\!\!\!\!\!\!\!\!\!\!\!\!\!\!\!\!\!\!\!\!\!\!\!\!\!
    -8.4687\E{-5 }x^{22/3}+2.14019\E{-5} x^8-1.16328\E{-6} x^{28/3}\!\!\!,
    \nonumber
\end{eqnarray}
%%%
obtained with a a more extended power-law expansion.
Instead, for $x>1$, we have started from a slightly more accurate asymptotic solution:
%%%
\begin{equation}
    F_{x\gg1,B}(x)\simeq\left(\frac{\upi}{2}x\right)^{1/2}e^{-x}\left(1+\frac{55}{72 x}\right),
\end{equation}
%%%
and we have then applied a correction of the form $1-\exp(f(\ln x))$ (where $\ln$ is the natural logarithm).
By performing a numerical fit we have then derived:
%%%
\begin{eqnarray}
    f(\ln x)\!\!\!\!&\simeq&\!\!\!\!-1.6144-0.997809
   \,\ln x-0.214177\,(\ln x)^2   \\   &&\!\!\!\!\!\!\!\!\!\!\!\!\!\!\!\!\!\!\!\!\!\!\!\!\!\!\!\!\!\!\!\!\!\!\!
   +0.00120319\,(\ln x)^3+0.00660814\,(\ln x)^4-0.000826152\,(\ln x)^5.
   \nonumber
\end{eqnarray}
%%%
In conclusion, an excellent fit to $F(x)$, everywhere better than about $10^{-6}$, and usually much better than this, is:
%%%
\begin{equation}
    F(x)=\left\{
    \begin{array}{ll}
     F_{x\ll1}(x)\cdot F_{\rm{corr,\,} x<1}(x)&\hbox{for\,}x<1  \\
     F_{x\gg1,B}(x)\cdot(1-\exp(f(\ln x))&\hbox{for\,}x\in[1.,720.]  \\
     F_{x\gg1,B}(x)&\hbox{for\,}x>720.
    \end{array}
    \right.
\end{equation}
%%%
In a similar way, for $G(x)$, we have found the $x<1$ correction:
%%%
\begin{eqnarray}
   G_{\rm{corr,\,} x<1}(x)&\simeq&\!\!\!\!\!\!
    1-1.17767 x^{4/3}+0.75 x^2-0.176651 x^{10/3}
    \nonumber\\ &&\!\!\!\!\!\!\!\!\!\!\!\!\!\!\!\!\!\!\!\!\!\!\!\!\!\!\!\!\!\!\!\!\!\!\!\!\!\!\!\!\!\!\!\!\!\!\!\!
    +0.0703125 x^4-0.0082805 x^{16/3}+0.00251116 x^6
    \\    &&\!\!\!\!\!\!\!\!\!\!\!\!\!\!\!\!\!\!\!\!\!\!\!\!\!\!\!\!\!\!\!\!\!\!\!\!\!\!\!\!\!\!\!\!\!\!\!\!
    -0.000188193 x^{22/3}+4.70843\E{-5} x^8-2.52045\E{-6} x^{28/3}\!\!\!,
    \nonumber
\end{eqnarray}
%%%
the more accurate asymptotic solution:
%%%
\begin{equation}
    G_{x\gg1,B}(x)\simeq\left(\frac{\upi}{2}x\right)^{1/2}e^{-x}\left(1+\frac{7}{72 x}\right),
\end{equation}
%%%
and the correction for $x>1$ based on the fitted function:
%%%
\begin{eqnarray}
    g(\ln x)\!\!\!\!&\simeq&\!\!\!\!-3.79105-1.52718
   \,\ln x-0.131875\,(\ln x)^2   \\   &&\!\!\!\!\!\!\!\!\!\!\!\!\!\!\!\!\!\!\!\!\!\!\!\!\!\!\!\!\!\!\!\!\!\!\!
   +0.0153602\,(\ln x)^3-0.000369143\,(\ln x)^4,
   \nonumber
\end{eqnarray}
%%%
so that $G(x)$ can be approximated, again to better than about $10^{-6}$, by using:
%%%
\begin{equation}
    G(x)=\left\{
    \begin{array}{ll}
     G_{x\ll1}(x)\,G_{\rm{corr,\,} x<1}(x)&\hbox{for\,}x<1  \\
     G_{x\gg1,B}(x),(1-\exp(g(\ln x))&\hbox{for\,}x\in[1.,720.]  \\
     G_{x\gg1,B}(x)&\hbox{for\,}x>720.
    \end{array}
    \right.
\end{equation}
%%%

%%%% Appendix B EXT %%%%%%%%%%%%%%%%%%%%%%%%%%%%%
\section{Fits to some integrals}
\label{app:fits2intgFG}
Let us introduce the quantities:
%%%
\begin{eqnarray}
    \Fn(x,s,\bt,\xcut)\!\!\!\!&=&\!\!\!\!\frac{s+1}{s+7/3}\,\Gm\left(\frac{s}{4}+\frac{7}{12}\right)^{-1}\Gm\left(\frac{s}{4}-\frac{1}{12}\right)^{-1}
    \nonumber\\
    &&\qquad F(x)\,\left(\frac{x}{2}\right)^{(s-3)/2}\exp\left(-\left(\frac{x}{\xcut}\right)^{-\bt/2}\right);\qquad\\
    \Gn(x,s,\bt,\xcut)\!\!\!\!&=&\!\!\!\!\qquad\quad\,\Gm\left(\frac{s}{4}+\frac{7}{12}\right)^{-1}\Gm\left(\frac{s}{4}-\frac{1}{12}\right)^{-1}
    \nonumber\\
    &&\qquad G(x)\,\left(\frac{x}{2}\right)^{(s-3)/2}\exp\left(-\left(\frac{x}{\xcut}\right)^{-\bt/2}\right).\quad
\end{eqnarray}
%%%
They are normalised in such a way that:
%%%
\begin{eqnarray}
    \int_0^\infty\Fn(x,s,\bt,0)\,dx&=&1;\\
    \int_0^\infty\Gn(x,s,\bt,0)\,dx&=&1.
\end{eqnarray}
%%%
The parameters $\xcut$ and $\bt$ correspond to those introduced in the main text, when we started considering a particle distribution with an energy cutoff (see Eq.~\ref{eq:ndistrib}).
Note that $\xcut=0$ corresponds to $\gmcut=\infty$, namely to the case of a power-law distribution, the case already treated in Paper~I.

Aim of this Appendix is to evaluate the integral quantities:
%%%
\begin{eqnarray}             
  \label{eq:XFdefapp}
  X_F(s,\bt,\xcut)&=&\int_0^\infty\Fn(x,s,\bt,\xcut)\,dx;\\
  \label{eq:XGdefapp}
  X_G(s,\bt,\xcut)&=&\int_0^\infty\Gn(x,s,\bt,\xcut)\,dx,
\end{eqnarray}
%%%
which are fundamental for the aims of the present work.

To rephrase in an explicit form what already present in Eqs.~\ref{eq:xdef}, \ref{eq:IPLprime}, \ref{eq:QPLprime} and \ref{eq:W0def}, the Stokes parameters in the case of a homogeneous MF and a particle energy distribution (Eq.~\ref{eq:ndistrib}) can be expressed as:
%%%
\begin{eqnarray}             
  {\cal I}&=&\frac{s+7/3}{s+1}\Gm\left(\frac{s}{4}+\frac{7}{12}\right)\Gm\left(\frac{s}{4}-\frac{1}{12}\right)\frac{A\,H}{4}\left(\frac{3e}{\elm c}\right)^{(s-1)/2}    \nonumber\\
  &&\quad\om^{-(s-1)/2}B^{(s+1)/2}X_F\left(s,\bt,\frac{2}{3}\frac{\elm c}{e}\frac{\om}{B\gmcut^2}\right);  \\
  {\cal Q}&=&\qquad\quad\Gm\left(\frac{s}{4}+\frac{7}{12}\right)\Gm\left(\frac{s}{4}-\frac{1}{12}\right)\frac{A\,H}{4}\left(\frac{3e}{\elm c}\right)^{(s-1)/2}\nonumber\\
  &&\quad\om^{-(s-1)/2}B^{(s+1)/2}X_G\left(s,\bt,\frac{2}{3}\frac{\elm c}{e}\frac{\om}{B\gmcut^2}\right);
\end{eqnarray}
%%%
Unfortunately, the integrals as in Eqs.~\ref{eq:XFdefapp} and \ref{eq:XGdefapp} do not seem to admit exact analytical solutions; so we have computed them numerically, in some specific cases, and then we have devised analytic approximations to them.

%%%%%%%%%%%%%%%%%%%%%%%%%%%%%%%%%%%%%%%%%%%%%%%%%
\begin{figure} 
\centering
	\includegraphics[width=.47\textwidth]{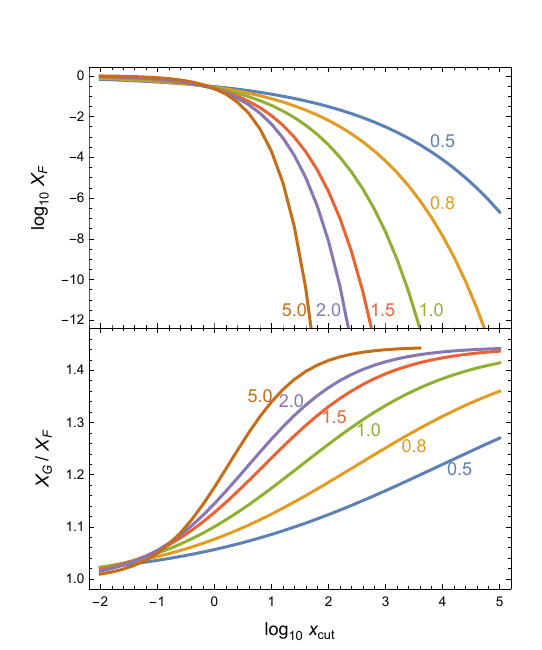}
        \caption{Plots of $\log_{10}X_F$ and of $X_G/X_F$, for the case $s=2$ and several values of $\bt$, equal to 0.5, 0.8, 1.0, 15, 2.0, 5.0. 
        }
\label{fig:figintgFandGovFglob}	
\end{figure}
%%%%%%%%%%%%%%%%%%%%%%%%%%%%%%%%%%%%%%%%%%%%%%%%%
\begin{figure} 
\centering
	\includegraphics[width=.47\textwidth]{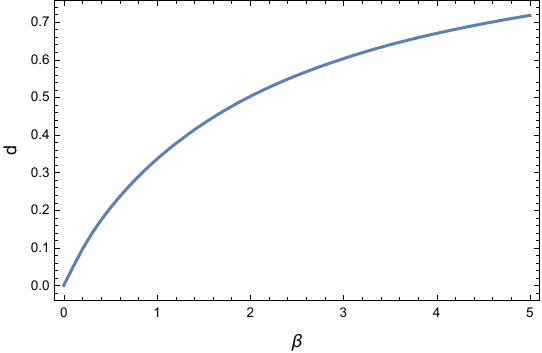}
        \caption{Dependence of the asymptotic exponential power ($d$) in $X_F$, as a function of $\bt$.
        }
\label{fig:figdofbeta}	
\end{figure}
%%%%%%%%%%%%%%%%%%%%%%%%%%%%%%%%%%%%%%%%%%%%%%%%%
As shown in Fig.~\ref{fig:figintgFandGovFglob}, the $X_F$ curves (upper panel) decrease very rapidly, at large $\xcut$ values; and this behaviour is more prominent for larger values of $\bt$.
The behaviour of $X_G$ is similar, so that their ratio (lower panel) is confined, between 1 and $(s+7/3)/(s+1)$, and the transition region is narrower for larger values of $\bt$.
Asymptotically for $\xcut\rightarrow\infty$, both $X_F$ and $X_G$ seem to vanish very rapidly.

For fitting their behaviour over the whole range of $\xcut$, we have chosen the following approximating formulae:
%%%
\begin{eqnarray}
    \label{eq:XFappr}
   \XFappr(s,\bt,\xcut)\!\!\!\!&=&\!\!\!\!\exp\left(-\,\frac{a\xcut^b}{(1+(a/c)^e\xcut^{e(b-d)})^{1/e}}\right);  \\
    \label{eq:XGappr}
    \frac{\XGappr(s,\bt,\xcut)}{\XFappr(s,\bt,\xcut)}\!\!\!\!&=&\!\!\!\!\frac{1+(s+7/3)/(s+1)\,(f+g\ln\xcut)\xcut^h}{1+(f+g\ln\xcut)\xcut^h}.\quad\quad
\end{eqnarray}
%%%
For construction, both formulae approach 1 for $\xcut\rightarrow1$, while $\XGappr/\XFappr$ approaches $(s+7/3)/(s+1)$ for large values of $\xcut$.

Tables~\ref{tab:XFpar} and \ref{tab:XGovFpar} list the best-fit coefficients, for a number of choices of $s$ and $\bt$.
The number of digits chosen for each coefficient is the minimum required to avoid the accuracy of the approximation to be appreciably downgraded by this truncation: in particular, the coefficients $b$ and $d$ really require a 7-digit accuracy.
The range of the $\xcut$ values for the numerical arrays, on which to perform the fits, has been chosen to be very wide, from $3\E{-7}$ to $3\E{6}$, with the exception of higher values of $\bt$, when the machine numerical underflow limits the extension of the upper boundary (by a factor $\sim5$ for $\bt=1.5$, or $\sim30$ for $\bt=2.0$).
In order to give an idea of the level of the accuracy reached, the last column of each table gives the maximum deviation of the approximation. The values in this column are even too conservative: for instance, in Table~\ref{tab:XFpar} typically the largest deviation occurs close to the upper limit of $\xcut$, where the value of $X_F$ is extremely low, and inessential for the calculation of other quantities.

An interesting dependence is how the asymptotic exponential power ($d$) for $X_F$ (and for $X_G$ as well), at large values of $\xcut$, changes with $\bt$: this dependence is shown in Fig.~\ref{fig:figdofbeta}.
First of of all, it is clear that it is not a linear dependence: for very low values of $\bt$, $d$ is well approximated by $\bt/2$; but then it soon flattens, so that for $\bt=5$ $d$ only reaches $0.7177$; due to numerical underflows at large values of $\bt$ and $\xcut$ (see above) its trend at larger values of $\bt$ it is not sufficiently reliable to verify if it saturates at a maximum value, and no analytical solutions have been found so far, even in these limiting cases.  
%%%
\begin{table}
\centering
\caption{Best-fit coefficients $(a,b,c,d,e)$, to adopt for the function $\XFappr(s,\bt,\xcut)$, in order to approximate the function $X_F(s,\bt,\xcut)$, for various choices of the parameters $s$ and $\bt$. The last column gives the maximum deviation, in absolute value, of the approximation.}
\label{tab:XFpar}
\setlength{\tabcolsep}{2pt}
\begin{tabular}{||c c c c c c c c||} 
 \hline
 $s$ & $\bt$ & $a$ & $b$ & $c$ & $d$ & $e$ & Max.Dev. \\ [0.5ex] 
 \hline\hline
 1.9 & 1.0 & 6.02774 & 0.5619608 & 1.90225 & 0.3330766 & 0.99241 & 0.0024 \\ 
 \hline
 1.9 & 1.5 & 7.11761 & 0.7251578 & 1.99772 & 0.4280908 & 0.95624 & 0.0062 \\ 
 \hline
 1.9 & 2.0 & 4.70687 & 0.7723972 & 2.02137 & 0.4993386 & 1.19300 & 0.0079 \\
 \hline
 2.0 & 1.0 & 5.01121 & 0.5586867 & 1.90537 & 0.3329990 & 1.01129 & 0.0019 \\ 
 \hline
 2.0 & 1.5 & 6.03545 & 0.7317618 & 1.99846 & 0.4280685 & 0.95430 & 0.0072 \\ 
 \hline
 2.0 & 2.0 & 4.42475 & 0.7957342 & 2.02102 & 0.4993466 & 1.13191 & 0.0083 \\
 \hline
 2.1 & 1.0 & 4.23132 & 0.5548417 & 1.90750 & 0.3329464 & 1.03592 & 0.0016 \\ 
 \hline
 2.1 & 1.5 & 5.17683 & 0.7359420 & 1.99879 & 0.4280585 & 0.95770 & 0.0069 \\ 
 \hline
 2.1 & 2.0 & 4.27174 & 0.8200959 & 2.02095 & 0.4993476 & 1.06808 & 0.0060 \\ [1.0ex] 
 \hline
\end{tabular}
\label{tab:fcoefs}
\end{table}
%%%

%%%
\begin{table}
\centering
\caption{Same as Table~\ref{tab:fcoefs}, but for $X_G(s,\bt,\xcut)/X_F(s,\bt,\xcut)$. Now the best-fit coefficients are $(f,g,h)$.}
\label{tab:XGovFpar}
\setlength{\tabcolsep}{2pt}
\begin{tabular}{||c c c c c c||} 
 \hline
 $s$ & $\bt$ & $f$ & $g$ & $h$ & Max.Dev. \\ [0.5ex] 
 \hline\hline
  \hline
 1.9 & 1.0 & 0.2995 & 0.0214 & 0.2784 & 0.0010 \\ 
 \hline
 1.9 & 1.5 & 0.4092 & -0.0212 & 0.4933 & 0.0028 \\ 
 \hline
 1.9 & 2.0 & 0.4913 & -0.0294 & 0.5704 & 0.0027 \\
 \hline
 2.0 & 1.0 & 0.2870 & 0.0217 & 0.2781 & 0.0009 \\ 
 \hline
 2.0 & 1.5 & 0.3912 & -0.0212 & 0.5005 & 0.0029 \\ 
 \hline
 2.0 & 2.0 & 0.4692 & -0.0297 & 0.5805 & 0.0029 \\
 \hline
 2.1 & 1.0 & 0.2754 & 0.0218 & 0.2780 & 0.0008 \\ 
 \hline
 2.1 & 1.5 & 0.3801 & 0.0376 & 0.3551 & 0.0010 \\ 
 \hline
 2.1 & 2.0 & 0.4556 & 0.0504 & 0.4136 & 0.0010 \\ [1ex] 
 \hline
\end{tabular}
\end{table}
%%%

%%%%%%%%%%%%%%%%%%%%%%%%%%%%%%%%%%%%%%%%%%%%%%%%%
\begin{figure} 
\centering
	\includegraphics[width=\columnwidth]{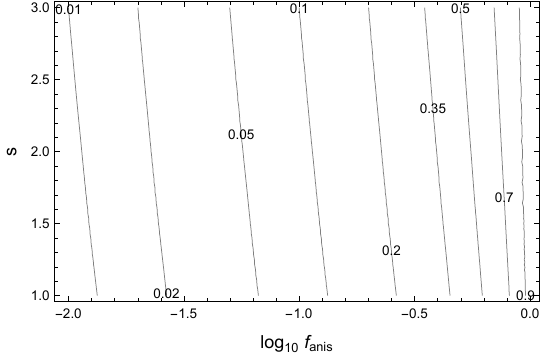}
        \caption{Map of $\Pi/\Pio$ for anisotropic fluctuations and a vanishing ordered MF (the correct version of Fig.~4 in Paper I).
        }
\label{fig:corrfig4}	
\end{figure}
%%%%%%%%%%%%%%%%%%%%%%%%%%%%%%%%%%%%%%%%%%%%%%%%%

%%%% Appendix C EXT %%%%%%%%%%%%%%%%%%%%%%%%%%%%%
\section{Errata-corrige to Paper I}

Let us take the opportunity, in this Appendix, to correct two errors present in Paper I.

First, a misprint in the second line of Eq.~21: instead of $(B^2-2B_y^2)/\sqrt{B_y^2}$ there should have been written $(2B_y^2-B^2)/\sqrt{B^2-B_y^2}$. However, we point out that the final formula, in the third line of that equation, is correct.

Then, Fig.~4: in spite of the labels on the axes, for this figure we had erroneously used the same data matrix as for the isotropic case. Its correct version is now shown in Fig.~\ref{fig:corrfig4}. One may notice that now, as also intuitively expected, the polarization degree increases with increasing the anisotropy parameters $\fanis$.

\end{appendix}

%
%TC:endignore
\end{document}